\documentclass[12pt]{article}
\usepackage{a4}
\usepackage{epsfig}

\newlength{\abstwidth}
\def\be{\begin{equation}} 
\def\ee{\end{equation}} 
\def\bq{\begin{equation}} 
\def\eq{\end{equation}} 
\def\bqa{\begin{eqnarray}} 
\def\eqa{\end{eqnarray}} 
\def\lsim{\roughly<}
\def\gsim{\roughly>}

\begin{document}

\setcounter{section}{0}

\renewcommand{\theequation}{%
\mbox{\arabic{section}.\arabic{equation}}}

\def\lsim{\mathrel{\rlap{\lower4pt\hbox{\hskip1pt$\sim$}}
    \raise1pt\hbox{$<$}}}         
\def\gsim{\mathrel{\rlap{\lower4pt\hbox{\hskip1pt$\sim$}}
    \raise1pt\hbox{$>$}}}         

\pagestyle{empty}

\begin{flushright}
BI-TP 2003/23\\
\end{flushright}

\vspace{\fill}

\begin{center}
{\Large\bf The total virtual photoabsorption cross section, deeply virtual
Compton scattering and vector-meson production$^*$}
\\[3.5ex]
{\bf Masaaki Kuroda$^{* \dagger}$ {\rm and} 
Dieter Schildknecht$^\dagger$ }\\[2.5mm]
$^*$ Institute of Physics, Meiji-Gakuin University \\[1.2mm]
Yokohama 244, Japan \\[1.2mm]
$^\dagger$Fakult\"{a}t f\"{u}r Physik, Universit\"{a}t Bielefeld \\[1.2mm] 
D-33501 Bielefeld, Germany \\[1.5ex]
\end{center}

\vspace{\fill}

\begin{center}
{\bf Abstract}\\[2ex]
\begin{minipage}{\abstwidth}
Based on the two-gluon-exchange dynamical mechanism for deeply inelastic
scattering at low $x \simeq Q^2/W^2 \ll 1$, we stress the intimate 
direct connection
between the total virtual photoabsorption cross section, deeply virtual
Compton scattering and vector-meson electroproduction. A simple expression
for the cross section for deeply virtual Compton scattering is derived. 
Parameter-free predictions are obtained for deeply-virtual Compton forward
scattering and vector-meson forward production, once the parameters in the total
virtual photoabsorption cross section are determined in a fit to the 
experimental data on deeply inelastic scattering. Our predictions are
compared with the experimental data from HERA.
\end{minipage}
\end{center}

\vspace{\fill}
\noindent

\rule{60mm}{0.4mm}\vspace{0.1mm}

\noindent
${}^*$ Supported by Deutsche Forschungsgemeinschaft, contract number schi 189/6-2.\\
\clearpage
\pagestyle{plain}
\setcounter{page}{1}

\section{Introduction}

We have recently stressed and worked out\cite{Ku-Schi, Schi-Ku} the intimate
connection between deeply inelastic scattering (DIS) at $x \simeq Q^2/W^2
\ll 1$, i.e. between the virtual photoabsorption cross section including
$Q^2 = 0$ photoproduction, and ``elastic'' diffractive production,
\be
\gamma^*p \to (q \bar q)^{J=1} p.
\label{(1)}
\ee
In (\ref{(1)}), $(q \bar q)^{J=1}$ may refer to one of the discrete
vector mesons, $\rho^0, \omega, \phi, J/\psi, \Upsilon$, or else, 
to the
diffractively produced mass continuum under the restriction to spin $J=1$.
Deeply virtual Compton scattering, $\gamma^* p \to \gamma p$, is to be
included in (\ref{(1)}) by attaching a real photon to the final 
$(q \bar q)^{J=1}$ state, via
\be
\gamma^*p \to [\int (q \bar q)^{J=1} \to \gamma] p ,
\label{1.2}
\ee
where the integration runs over the mass of the intermediate 
$(q \bar q)^{J=1}$ state.

The treatment in refs. \cite{Ku-Schi, Schi-Ku} was relying on a single 
assumption: it is the generic two-gluon-exchange structure\cite{Low, NZ}
depicted in figs. 1 and 2, that is the basic dynamical mechanism underlying
the total photoabsorption cross section and diffractive production
at low $x$.

In the present work we will employ the connection between the total
photoabsorption cross section and diffractive production to obtain 
parameter-free predictions for deeply virtual Compton scattering,
$\gamma^*p\to \gamma p$ (DVCS),
including real Compton scattering, $\gamma p \to \gamma p$, and 
for forward vector-meson
electroproduction, including vector-meson photoproduction. 
Both, DVCS and vector-meson production, will be compared
with the experimental data from HERA.

Section 2 summarizes the salient features of our approach to the total
virtual photoabsorption cross section and diffractive production. 

In Section 3, on the basis of Section 2, we will give a simple derivation
of an extraordinarily simple expression for DVCS that will be shown to
agree with the data \cite{HERA-I} from HERA with respect to the 
$W$ and the $Q^2$
dependence.

In Section 4, we turn to vector-meson production and compare with the
experimental data \cite{HERA-II} with respect to the $W$ and the $Q^2$ 
dependence and the
longitudinal-to-transverse ratio.

We end with brief conclusions in Section 5.

\section{The total virtual photoabsorption cross section}
\setcounter{equation}{0}

The intimate connection between the total photoabsorption cross section and
``elastic'' diffraction is explicitly represented by the recently
derived sum rules\cite{Ku-Schi, Schi-Ku}

\be
\sigma_{\gamma^*_T p} (W^2 , Q^2) = \sqrt{16\pi} 
\sqrt{\frac{\alpha R_{e^+ e^-}}{3\pi}} \int_{m^2_0} dM^2 \frac{M}{Q^2 + M^2} 
\sqrt{\left.\frac{d\sigma_{\gamma^*_T p
\rightarrow (q \bar q)^{J=1}_T p}}{dt dM^2}\right|_{t=0}} 
\label{2.1} 
\ee
and
\be
\sigma_{\gamma^*_L p} (W^2 , Q^2) = \sqrt{16\pi} 
\sqrt{\frac{\alpha R_{e^+ e^-}}{3\pi}} \int_{m^2_0} dM^2 \frac{\sqrt{Q^2}}
{Q^2 + M^2} 
\sqrt{\left.\frac{d\sigma_{\gamma^*_L p
\rightarrow (q \bar q)^{J=1}_L p}}{dt dM^2}\right|_{t=0}} .
\label{2.2} 
\ee
that relate the transverse and the longitudinal  part of the total cross
section to the transverse and the longitudinal part, respectively, of the 
forward-production amplitude of $(q \bar q)^{J=1}$ (vector) states. The
representations (\ref{2.1}) and (\ref{2.2}) follow from the 
assumption that the two-gluon-exchange dynamical mechanism of figs. 1
and 2 be valid in the $x \to 0$ limit. The imaginary part of the
$(q \bar q)^{J=1}$-forward-production amplitude in (\ref{2.1}) and 
(\ref{2.2}) appears as the square root of the forward-production cross
section\footnote{Strictly speaking, forward production of (massive)
$(q \bar q)$ vector states involves a finite momentum transfer, 
$\vert t_{min} \vert$, and the sum rules (\ref{2.1}) and (\ref{2.2}) require
an extrapolation to $t = 0$. We will employ the approximation of
$\vert t_{min} \vert \simeq 0$ for diffractive forward production, since
$\vert t_{min} \vert$ becomes exceedingly small for $x \to 0$. Compare the
explicit formula for $\vert t_{min} \vert$ in Section 4.}
\footnote{In (\ref{2.1}) and (\ref{2.2}), we have suppressed
factors $1/\sqrt{(1 + \beta^2_T)}$ and $1/\sqrt{(1 + \beta^2_L)}$,
respectively, where $\beta_{T,L}$ stands for the ratio of the 
real-to-imaginary part of the forward scattering amplitude. The ratio
$\beta_{T,L}$ is expected to be small, $\beta_{T,L} \ll 1$, as in truly elastic
hadron-hadron scattering, and $\beta^2_{T,L}$ may frequently be ignored.}
\be
\frac{d \sigma_{\gamma^*_{T,L}p \to (q \bar q)^{J=1}_{T,L}p}}{dtdM^2}
\left|_{t = 0} \right. = 
\frac{d \sigma_{\gamma^*_{T,L}p \to (q \bar q)^{J=1}_{T,L}p}}{dtdM^2}
\left|_{t = 0} \right. (W^2, Q^2, M^2).
\label{2.3}
\ee
As indicated in (\ref{2.3}), the 
forward-production cross section depends on the total $\gamma^* p$ 
center-of-mass energy, $W$, and the photon four-momentum squared, 
$q^2 = - Q^2$, as well as the mass $M \equiv M_{q \bar q}$ of the
$(q \bar q)$ vector state being produced and integrated over in (\ref{2.1})
and (\ref{2.2}). The lower limit, $m^2_0$, in (\ref{2.1})
and (\ref{2.2}), with $m_0$ being smaller than the $\rho^0$-meson mass,
$m^2_0 < m^2_{\rho^0}$, enters via quark-hadron duality\cite{Sakurai, 
Devenish,Levin}; 
the
vector mesons, $\rho^0, \omega, \phi, J/\psi$ and $\Upsilon$ are 
treated as part
of the diffractive continuum rather than being added as separate
resonances. Indeed, photoabsorption for spacelike photons should be 
insensitive to fine details of the photon coupling to $q \bar q$ pairs
in the timelike region and a global description in which a continuous
spectral weight function interpolates\cite{Devenish} 
the low-lying vector-meson
resonances is expected to be successful. The sum of the 
squares of the active quark charges (in units of $e$) in (\ref{2.1})
and (\ref{2.2}) is expressed by
\be
R_{e^+e^-} = 3 \sum_q Q^2_q,
\label{2.4}
\ee
where $R_{e^+e^-}$ denotes the cross section for $e^+ e^-$ annihilation,
$R_{e^+e^-} \to (q \bar q)^{J=1} \to$ hadrons, in units of the cross
section for $e^+e^- \to \mu^+\mu^-$.

The sum rules (\ref{2.1}) and (\ref{2.2}) were derived \cite{Ku-Schi, Schi-Ku}
by comparing with each other the explicit expressions for the total cross 
section and the cross section for diffractive production based on 
the two-gluon-exchange
interaction depicted in figs. 1 and 2 in the limit of $x \simeq Q^2/W^2 \ll 1$.
In the frequently employed transverse-position-space representation 
\cite{Bjorken,NZ} the total
cross section and the cross section for diffractive production of 
$(q \bar q)^{J=1}$ (vector) states are then given by \cite{Schi-Ku, Ku-Schi}
\begin{eqnarray}
& & \sigma_{\gamma^*_{T,L}p} (W^2 , Q^2) = \label{2.5} \\
& & = \int dz \int d^2 r_\bot 
| \psi_{T,L} (r_\bot , z(1-z), Q^2) |^2 
\sigma_{(q \bar q)^{J=1}_{T,L}p}
(\vec r_\bot \sqrt{z(1-z)}, W^2) 
\nonumber
\end{eqnarray} 
and
\begin{eqnarray}
& & \frac{d\sigma_{\gamma^*_{T,L}p \to (q \bar q)^{J=1}_{T,L}}}{dt}
\left|_{t=o} \right.
= \label{2.6} \\
& & = \frac{1}{16 \pi} \int dz \int d^2 r_\bot \left| \psi_{T,L}
(r_\bot , z (1-z), Q^2) \right|^2
\sigma^2_{(q \bar q)^{J=1}_{T,L} p}
(r_\bot \sqrt{z (1-z)}, W^2) ,
\nonumber
\end{eqnarray}
where
\be
\sigma_{(q \bar q)^{J=1}_{T,L}} (\vec r_\bot \sqrt{z(1-z)}, W^2) = \int d^2 l^
\prime_\bot \bar\sigma_{(q \bar q)^{J=1}_{T,L}p} (\vec l^{~\prime 2}_\bot , 
W^2)(1 - e^{-\vec l^{~\prime}_\bot \cdot \vec r_\bot \sqrt{z(1-z)}}) .
\label{2.7}
\ee
In (\ref{2.5}), the restriction that the photon exclusively couples to
spin $J=1$ quark-antiquark pairs, $(q \bar q)^{J=1}$, is explicitly
incorporated. The dependence of the color-dipole cross section,
$\sigma_{(q \bar q)^{J=1}_{T,L}} (r_\bot \sqrt{z (1-z)}, W^2)$
on $r_\bot \sqrt{z (1-z)}$ assures the restriction to the scattering
of $J = 1$ color dipoles on the proton. As usual, in (\ref{2.5}) and
(\ref{2.6}) the (light-cone) wave function of the virtual photon is
denoted by $\psi_{T,L} (r_\bot, z (1-z), Q^2)$, where
\bqa
  \Big| \psi_T (r_\bot , z (1-z), Q^2) \Big|^2
  &=& {{6\alpha}\over{(2\pi)^2}}\sum_q Q_q^2\Bigl[
     \Bigl(z^2+(1-z)^2\Bigr)\epsilon^2K_1(\epsilon r_\perp)^2
     \nonumber \\
  & & ~~~~~~~~~~~~~~~~~~~
     +m_q^2K_0(\epsilon r_\perp)^2\Bigr], \label{2.8} \\
 \left| \psi_L (r_\bot , z (1-z), Q^2) \right|^2
  &=& {{6\alpha}\over{(2\pi)^2}}\sum_q Q_q^2\Bigl[
     4Q^2z^2(1-z)^2K_0(\epsilon r_\perp)^2\Bigr], \label{2.9} \\
      \epsilon^2 &=& z(1-z)Q^2+m_q^2. \label{2.10}
\eqa
In standard notation, $Q_q$ and $m_q$ denote the quark charge and mass, 
respectively, and $z$ is the fraction of the light-cone momentum carried by the
quark from the $q \bar q$ pair, i.e. $k_+ = z q_+ = z (q^0 + q^3)$. 
Finally, $K_1$ and $K_0$ denote modified Bessel functions. 

Upon inserting (\ref{2.7}) into (\ref{2.5}) and (\ref{2.6}) and 
introducing the mass of the $(q \bar q)^{J=1}$ state, $M \equiv
M_{(q \bar q)^{J=1}}$, and upon reducing the number of integrations, one
reads off the validity of the sum rules (\ref{2.1}) and (\ref{2.2}).
\footnote{ The approximation of massless quarks was employed throughout.}

The representations (\ref{2.5}) to (\ref{2.7}) differ from the 
frequently employed ones\cite{NZ}
\be
\sigma_{\gamma^*_{T,L} p} (W^2 , Q^2 ) = \int dz \int d^2 r_\bot \left| 
\psi_{T,L} (r_\bot , z (1-z) , Q^2 ) \right|^2 
\sigma_{(q \bar q )p} (\vec r_\bot , W^2 ), 
\label{2.11}
\ee 
and
\begin{eqnarray}
\left.\frac{d\sigma_{\gamma^*_{T,L} p \rightarrow Xp} (W^2 , Q^2)}{dt}  
\right|_{t=0}
  &=& \frac{1}{16\pi} \int dz \int d^2 r_\bot \left| \psi_{T,L} 
(r_\bot , z(1-z) , 
    Q^2 ) \right|^2  \nonumber \\ 
& &~~~~~~~~~  \sigma^2_{(q \bar q )p} (\vec r_\bot , W^2 ) ,
\label{2.12}
\end{eqnarray}
with
\be
\sigma_{(q \bar q)p} (r_\bot , W^2) = 
\int d^2 l_\bot \tilde\sigma_{(q \bar q)p} \left( \vec l^{~2}_\bot ,
W^2 \right) \left( 1 - e^{-i \vec l_\bot \cdot \vec r_\bot} \right)
\label{2.13}
\ee
by the substitutions
\begin{eqnarray}
\vec r_\bot &\to & \vec r^{~\prime}_\bot = \vec r_\bot \sqrt{z (1-z)}, 
\nonumber \\
\vec l_\bot &\to & \vec l^{~\prime}_\bot = \frac{\vec l_\bot}{\sqrt{z (1-z)}},
\label{2.14}
\end{eqnarray}
and
\be
z (1-z) \tilde \sigma_{(q \bar q)p} (\vec l_\bot^{~\prime 2} z (1-z), W^2)
\to \bar \sigma_{(q \bar q)^{J=1}_{T,L}p} (\vec l_\bot^{~\prime 2}, W^2).
\label{2.15}
\ee
The replacement defined by (\ref{2.14}) and (\ref{2.15}) is a consequence
of the partial-wave decomposition of the $q \bar q$ states. The partial-wave
decomposition explicitly eliminates all contributions of $(q \bar q)^{J \not=
1}$ states that are irrelevant in (\ref{2.11}), since they are projected to
zero by the square of the photon wave function. The elimination of these
redundant components of the dipole cross section is of particular
importance, if the experimental data on $\sigma_{\gamma^*p} (W^2, Q^2)$
are used to extract the dipole cross section by employing a fitting
procedure.

A fit based on (\ref{2.5}) yields a unique prediction, according to 
(\ref{2.6}), for the (whole) diffractive production of 
$(q \bar q)^{J=1}_{T,L}$ states. A fit based on (\ref{2.11}), however,
does {\it not} yield a prediction for diffractive production according
to (\ref{2.12}). The contributions from diffractively produced states
$(q \bar q)^{J \not= 1}$, necessarily included in (\ref{2.12}), remain
unconstrained in a fit based on (\ref{2.11}). The fact that various
different dipole cross sections \cite{Landshoff} lead to equally good
representations of $\sigma_{\gamma^*p}(W^2, Q^2)$ is presumably a consequence
of the presence of such $J \not= 1$ contributions that remained
undetermined in the fits to $\sigma_{\gamma^*p} (W^2, Q^2)$. The
representation (\ref{2.11}) by itself is highly non-unique with respect
to the color-dipole cross section. The alternative of a simultaneous fit
of (\ref{2.11}) and (\ref{2.12}) would have to cope with the much larger 
experimental uncertainty of the experimental data on diffractive production
in comparison with the ones on $\sigma_{\gamma^* p} (W^2, Q^2)$. It is
preferable, accordingly, to use  the representation
(\ref{2.5}) with (\ref{2.7}) for a fit, and turn to a description of the total
diffractive production that includes $(q \bar q)^{J \not= 1}$
states subsequently.

Coming back to the sum rules (\ref{2.1}) and (\ref{2.2}) for a moment, we note
that they can be written down directly, without referring to the 
two-gluon-exchange dynamics, provided one is willing to adopt the validity
of generalized vector dominance (GVD, \cite{Sa-Schi,Devenish}), 
and (\ref{2.1}) was
indeed first given from GVD \cite{Spiesberger}. In terms of GVD, the factor
\be
\sqrt{\frac{\alpha R_{e^+e^-}}{3 \pi} \frac{1}{M^2}} \frac{M^2}{Q^2+M^2}
\label{2.16}
\ee
in (\ref{2.1}) originates from the coupling of the photon to the
$(q \bar q)^{J=1}$ intermediate state in fig. 2 and its propagation
(always assuming $x \to 0$), once a virtual photon, $\gamma^*_T$, is attached
to it, in order to reproduce the corresponding diagram of fig. 1. 
Compare fig. 3. In the case
of the longitudinal photon, $\gamma^*_L$, in (\ref{2.2}), an additional
factor $\sqrt{Q^2/M^2}$ must be included in GVD\cite{Fraas,Sa-Schi}
as a consequence of the coupling
of the photon to a conserved source.

   An additional general comment may be appropriate at this point 
concerning the emergence of GVD as a consequence of the
two-gluon-exchange
dynamical mechanism of fig.1 in the $x\to 0$ limit.
The essential point is the observation, new by no means,
\cite{css1, NZ, frankfurt} that
the denominators of the quark propagators in fig.1 in the 
$x\to 0$ limit in the photoabsorption cross section 
lead to factors of the form
\bq
     {1\over {Q^2+M^2}}, \label{2.17}
\eq
with
\bq
     M^2 = M_{q\bar q}^2 = {{\vec k^2_\perp+m_q^2}\over{z(1-z)}}.
   \label{2.18}
\eq
In (\ref{2.17}), one recognizes the denominator of a vector-meson 
propagator for a vector-meson  ($q\bar q$) state of mass $M$.  
Note that the propagator mass $M_{q \bar q}$ in (\ref{2.18}) is related
to the spacelike
four momentum squared $q^2$ of the photon the $q \bar q$ 
pair is coupled to,
\bq
   q^2 = {{k_q^2+\vec k_\perp^2}\over z} 
       + {{k_{\bar q}^2+\vec k_\perp^2}\over {1-z}} =-Q^2<0,
   \label{2.19}
\eq
by substitution of the on-shell values 
\bq
    k_q^2=k_{\bar q}^2=m_q^2,   \label{2.20} 
\eq
for both the quark and the antiquark four momentum.
The $q \bar q$ mass in the propagator, i.e. the mass of the 
$q \bar q$ state the photon dissociates or "fluctuates" into,
is accordingly identical to the mass of a $q \bar q$ state (with
on-shell quarks)
diffractively produced via two-gluon exchange.  In short, the quark
propagator becomes transmogrified into a vector-meson propagator 
containing  the invariant mass (\ref{2.18}) of a system of an on-shell 
quark and an 
on-shell antiquark, i.e. we have the GDV structure.

So far, no specific realization of the lower vertex in figs. 1 and 2 was
adopted. Examining the dipole cross section (\ref{2.7}) in the limit of
large and small interquark separation, $r^\prime_\bot \to \infty$ and
$r^\prime_\bot \to 0$, allows one to strongly reduce arbitrariness. From
(\ref{2.7})\footnote{It is to be stressed that ``saturation'' for
$r^\prime_\bot \to \infty$ and ``color transparency'' for
$r^\prime_\bot \to 0$ in (\ref{2.21}) are a consequence of the 
QCD-gauge-theory structure that is contained in (\ref{2.7}). 
Compare  ref. \cite{NZ}.}
\footnote{We drop the indices $T,L$, anticipating the ansatz
(\ref{2.24}) that does not contain a dependence on whether the 
$(q \bar q)^{J=1}$ state is transversely or longitudinally polarized.},
\be
\sigma_{(q \bar q)^{J=1}p} (r^\prime_\bot, W^2) = \sigma^{(\infty)} (W^2)
\cdot \left\{ \begin{array}{r@{\quad,\quad}l}
 1 &{\rm for~} r^\prime_\bot \to \infty,\\
\frac{1}{4} r^{\prime 2}_\bot \langle \vec l^{~\prime 2}_\bot \rangle_{W^2}
&{\rm for~} r^{\prime 2}_\bot \langle \vec l^{\prime 2}_\bot \rangle_{W^2}\to 0,
\end{array} \right. 
\label{2.21}
\ee
where
\be
\sigma^{(\infty)} (W^2) = \pi \int d \vec l^{~\prime 2}_\bot
\bar \sigma_{(q \bar q)^{J=1}} (\vec l^{~\prime 2}_\bot, W^2),
\label{2.22}
\ee
and
\be
\langle \vec l^{~\prime 2}_\bot \rangle_{W^2} =
\frac{\int d \vec l^{\prime 2}_\bot \vec l^{~\prime 2}_\bot 
\bar \sigma_{(q \bar q)^{J=1}} (\vec l^{~\prime 2}_\bot, W^2)}
{\int d \vec l^{\prime 2}_\bot
\bar \sigma_{(q \bar q)^{J=1}} (\vec l^{~\prime 2}_\bot, W^2)}
\equiv \Lambda^2 (W^2),
\label{2.23}
\ee
we find that the $J=1$ color-dipole cross section is strongly
constrained, once the integral (\ref{2.22}) over the transverse-gluon-momentum
dependence and its first moment, (\ref{2.23}), are specified. Even though
given values of the energy-dependent quantities $\sigma^{(\infty)} (W^2)$
and $\langle \vec l^{~\prime 2}_\bot \rangle_{W^2} \equiv \Lambda^2 (W^2)$
do by no means uniquely specify the color-dipole cross section, different
functional forms of $\sigma_{(q \bar q)^{J=1}p} (r^\prime_\bot, W^2)$
restricted by identical integrated distributions (\ref{2.22}) and
(\ref{2.23}) are nevertheless largely equivalent. According to (\ref{2.21})
they lead to identical color-dipole cross sections in both the limit of
large and the limit of small interquark separation. 

The ansatz for the
$J=1$ color-dipole cross section chosen previously (generalized vector 
dominance-color-dipole picture, GVD-CDP)
\cite{Cvetic}
is parametrized in terms of $\sigma^{(\infty)} (W^2)$ and
$\Lambda^2 (W^2)$,
\be 
\bar\sigma_{(q \bar q)^{J=1}_T} (\vec l^{~\prime 2}_\bot , W^2) = 
\bar\sigma_{(q \bar q)^{J=1}_L} (
\vec l^{~\prime 2}_\bot , W^2) = \sigma^{(\infty)} 
(W^2) \cdot \frac{1}{\pi} \delta
(\vec l^{~\prime 2}_\bot - \Lambda^2 (W^2)) , 
\label{2.24}
\ee 
and it can indeed be adopted without significant loss of generality. In
other words, the ansatz (\ref{2.24}) that parametrizes the $J=1$ color-dipole
cross section in terms of the two W-dependent parameters (\ref{2.22})
and (\ref{2.23}) is a fairly stringent consequence from the underlying
two-gluon-exchange mechanism. The specific choice of the $\delta$-function
in (\ref{2.24}) is a purely technical simplification.

With respect to transverse position space, according to (\ref{2.7}),
(\ref{2.24}) becomes
\be
\sigma_{(q \bar q)^{J=1}p} (\vec r^{~\prime 2}_\bot, W^2) = \sigma^{(\infty)}
(W^2) (1 - J_0 (r^\prime_\bot \Lambda (W^2))),
\label{2.25}
\ee
where $J_0 (r^\prime_\bot \Lambda (W^2))$ denotes a Bessel function.

A general conclusion on the energy dependence of the total cross section
follows immediately upon substituting (\ref{2.21}) into (\ref{2.7}) and
subsequently (\ref{2.7}) into the expression for the total cross section
(\ref{2.5}). Provided we exclude the (artificial) assumption that the first
moment of the transverse-momentum distribution (\ref{2.23}) be independent
of $W$, i.e. provided we exclude the assumption of $\Lambda^2 (W^2) =
{\rm const.}$, the increasing importance in (\ref{2.5})
of the short-distance limit
(\ref{2.21}) with increasing $Q^2$ will imply an increasingly stronger
dependence of $\sigma_{\gamma^*p} (W^2, Q^2)$ on $W$. This is indeed,
what is observed experimentally\cite{HERA-III}.

A theoretical argument against $\Lambda^2 (W^2) = {\rm constant}$ may
also be based on requiring duality between the two-gluon-exchange
dynamics and Regge (or more specifically Pomeron) behavior in the
$Q^2 = 0$ photoproduction limit. Requiring the energy dependence
in photoproduction from (\ref{2.5}) and (\ref{2.25}) to coincide with the
one due to Pomeron exchange,
\begin{eqnarray}
&\sigma^{Pomeron}_{\gamma p} (W^2) \equiv
\sigma ^{(\infty)} (W^2) \int dz
\int d^2 r_\bot \bigg| \psi_T (r_\bot, z (1-z), Q^2=0) \bigg|^2  \cdot
\nonumber \\
& (1 -J_0 (r_\bot \sqrt{z (1-z)} \Lambda (W^2))),
\label{2.26}
\end{eqnarray}
and to
be a genuine consequence of the generic two-gluon exchange 
structure\footnote{
The smooth transition to photoproduction may also be used as an argument
for the dipole cross section in (\ref{2.21}) to (\ref{2.23}) 
to depend on the single variable
$W^2 = x/Q^2$ rather than on $x$, or on both $x$ and $Q^2$ independently.
Note that in addition to the propagator factor (\ref{2.17}), it is
the dependence of the dipole cross section on the single
variable $W$ that fully guarantees the GVD structure.}, 
a factorized $W$ dependence, unrelated to the structure
of the two-gluon exchange dynamics that is contained in the integral in 
(\ref{2.26}), is theoretically disfavored. In other words, the option
$\Lambda = {\rm constant}$ that imposes soft-Pomeron behavior by an
overall factor in (\ref{2.26}) may safely be dismissed on theoretical
grounds.

The theoretical argument is supported by a fit \cite{Cvetic}
to the experimental data \cite{HERA-III}
that uses the parameterization of $\sigma_{\gamma p} (W^2)$ 
by Pomeron exchange as an
input at $Q^2=0$, thus replacing $\sigma^{(\infty)} (W^2)$ by 
$\sigma^{Pomeron}_{\gamma p}$ according to (\ref{2.26}). The fit  to the
experimental data indeed gave an increase of $\Lambda^2 (W^2)$ with
$W^2$, while $\sigma^{(\infty)}$ turned out to be energy independent
in good approximation.

In momentum space, the ansatz (\ref{2.24}), (\ref{2.25}) for the total 
cross section 
(\ref{2.5}) implies \cite{Ku-Schi}
\begin{eqnarray}
& & \sigma_{\gamma^*_T p} (W^2 , Q^2) = 
\frac{\alpha R_{e^+ e^-}}{3\pi} \sigma^{(\infty)} 
\int_{m^2_0} d M^2 \frac{1}{Q^2 + M^2}\nonumber \\ 
& & \cdot
\left[ \frac{M^2}{Q^2 + M^2} - \frac{1}{2} \left( 1 + \frac{M^2 - \Lambda^2 
(W^2) - Q^2}{\sqrt{(Q^2 + M^2 - \Lambda^2 (W^2))^2 + 4 Q^2 \Lambda^2 (W^2)}} 
\right) \right]
\label{2.27}
\end{eqnarray}
and
\begin{eqnarray}
& & \sigma_{\gamma^*_L p} (W^2 , Q^2) =
\frac{\alpha R_{e^+ e^-}}{3\pi} \sigma^{(\infty)} 
\int_{m^2_0} d M^2 \frac{1}{Q^2 + M^2} \nonumber \\
& & \cdot \left[ \frac{Q^2}{Q^2 + M^2}- \frac{Q^2}
{\sqrt{(Q^2 + M^2 - \Lambda^2 (W^2)) + 4 Q^2 \Lambda^2 (W^2)) 
}} 
\right].
\label{2.28}
\end{eqnarray}
The result of the integration of (\ref{2.27}) and (\ref{2.28}) was given
before \cite{Ku-Schi}. Here, in connection with the procedure 
used in Sections 3 and 4, it
is useful to note that the $Q^2 \to 0$ and the $Q^2 \gg \Lambda^2 (W^2)$ limit
of $\sigma_{\gamma^*_{T,L}p} (W^2, Q^2)$ can immediately be derived by
taking the corresponding limits under the integrals in (\ref{2.27}) and
(\ref{2.28}). For $Q^2 \ll \Lambda^2 (W^2)$, or, equivalently,
$\eta \ll 1$, one finds
\be
\sigma_{\gamma^*p} (W^2,Q^2) = \frac{\alpha}{3 \pi} R_{e^+e^-} 
\sigma^{(\infty)}
\ln \frac{1}{\eta},
\label{2.29}
\ee
while for $Q^2 \gg \Lambda^2 (W^2)$, or $\eta \gg 1$,
\be
\sigma_{\gamma^*_T p} (W^2, Q^2) = 2 \sigma_{\gamma^*_L p} (W^2, Q^2) =
\frac{\alpha}{3 \pi} R_{e^+e^-} \sigma^{(\infty)} \frac{1}{3 \eta},
\label{2.30}
\ee
with the scaling variable \cite{Diff2000}
\be
\eta (W^2, Q^2) = \frac{Q^2 + m^2_0}{\Lambda^2 (W^2)}.
\label{2.31}
\ee
The fit to the experimental data with the ansatz (\ref{2.24})
gave\cite{Surrow}\footnote{We note that the energy scale $W_0$ may be
eliminated in favor of $Q^2\equiv Q_0^2=1$ GeV$^2$ and
the value of $x$ corresponding to $W_0^2$ and $Q_0^2$,
i.e. $x_0\equiv Q_0^2/W_0^2$.  We have in very good approximation
$$\Lambda^2(W^2)\cong B x_0^{C_2} \Bigl( {{W^2}\over{1 {\rm GeV}}}
\Bigr)^{C_2}= 0.340 \Bigl( {{W^2}\over{1 {\rm GeV}}}\Bigr)^{0.27}.$$}
\begin{eqnarray}
& \Lambda^2 (W^2) = \left\{ \matrix{&B (\frac{W^2}{W^2_0} + 1)^{C_2} , \cr
   &C^\prime_1 \ln \left(\frac{W^2}{W^{\prime 2}_0} + C^\prime_2 \right) \cr} 
\right.\nonumber \\
& m^2_0 = 0.15 \pm 0.04 GeV^2
\label{2.32}
\end{eqnarray}
where 
\begin{eqnarray}
B & = & 2.24 \pm 0.43 GeV^2, \nonumber \\
W^2_0 & = & 1081 \pm 124 GeV^2, \label{2.33} \\
C_2 & = & 0.27 \pm 0.01. \nonumber
\end{eqnarray}

   We emphasize the emergence of a "soft" energy dependence 
in (\ref{2.29}) and a "hard" one in (\ref{2.30}) as a 
strict consequence of the two-gluon-exchange dynamics
together with the natural assumption that the 
effective gluon transverse momentum (\ref{2.23}) is to increasee
with energy $W$.

Note that the normalization of the cross sections 
(\ref{2.27}) to (\ref{2.30}) is determined by 
the product $R_{e^+e^-}\sigma^{(\infty)}$. 
With $R_{e^+e^-} = 2$ for three
active flavors, relevant for photoproduction, 
the experimental data require\cite{Cvetic} 
\be
\sigma^{(\infty)} = 30mb \simeq 77.04 GeV^{-2}.
\label{2.34}
\ee

With respect to the treatment of vector-meson production in Section 3,
we also note the cross section for diffractive production. With the
ansatz (\ref{2.21}) and (\ref{2.6}), one obtains \cite{Ku-Schi}
\bqa
& & \frac{d\sigma_{\gamma^*_T p\rightarrow (q \bar q)^{J=1}_T}}
    {dtdM^2dz} \Bigg|_{t=0} = 
\frac{\alpha \cdot R_{e^+ e^-}}{3\cdot 16\pi^2} (\sigma^{(\infty)})^2 
    {3\over 2}\Bigl( z^2+(1-z)^2\Bigr){1\over{M^2}} \cdot \label{2.35}  \\
& & \cdot \left[ \frac{M^2}{Q^2 + M^2} - \frac{1}{2} \left( 1 + 
\frac{M^2 - \Lambda^2 (W^2) - Q^2}{\sqrt{(Q^2 + M^2 - \Lambda^2 (W^2))^2
+ 4 Q^2 \Lambda^2 (W^2)}} \right) \right]^2 , \nonumber
\eqa
and
\begin{eqnarray}
& & \frac{d\sigma_{\gamma^*_L p\rightarrow (q \bar q)^{J=1}_L}}
   {dtdM^2dz} \Bigg|_{t=0} = 
\frac{\alpha \cdot R_{e^+ e^-}}{3\cdot 16\pi^2} (\sigma^{(\infty)})^2 \cdot 
   6Q^2z(1-z) \label{2.36} \\
  & & \cdot\left[ \frac{1}{Q^2 + M^2} - 
\frac{1}{\sqrt{(Q^2 + M^2 - \Lambda^2 (W^2))^2 +
4 Q^2 \Lambda^2 (W^2)}} \right]^2, \nonumber
\end{eqnarray} 
and, finally, for the sum of (\ref{2.35}) and (\ref{2.36}),
\
\begin{eqnarray}
& & \frac{d\sigma_{\gamma^* p \rightarrow (q \bar q)^{J=1}p}}
     {dtdM^2} 
\Bigg|_{t=0} =   
\frac{\alpha R_{e^+ e^-}}{3\cdot 32\pi^2}(\sigma^{(\infty)})^2 \cdot
\label{2.37} \\
& & \cdot  {1\over{M^2}} \left[ 1 - 
\frac{(M^2 + Q^2)^2 - (M^2 - Q^2) \Lambda^2 (W^2)}{(M^2 + Q^2) 
\sqrt{(Q^2 + M^2 - \Lambda^2 (W^2))^2 + 4 Q^2 \Lambda^2 (W^2)}} 
\right] . \nonumber
\end{eqnarray}
Comparing (\ref{2.35}) and (\ref{2.36}) with (\ref{2.27}) and (\ref{2.28})
once again takes us back to the sum rules (\ref{2.1}) and (\ref{2.2}), now
based on the specification (\ref{2.21}) of the $J = 1$ color-dipole cross
section.

From (\ref{2.29}), for any fixed $Q^2$, for $W \to \infty$, we reach the 
photoproduction limit \cite{Surrow},
\be
\lim_{W^2 \to \infty \atop{Q^2 = {\rm const}}} \frac{\sigma_{\gamma^*p}
(W^2, Q^2)}{\sigma_{\gamma p} (W^2)} = 1.
\label{2.38}
\ee

It has been the aim of this Section 2 to prepare the ground for the
discussions on DVCS and on ``elastic'' diffraction, in particular
vector-meson production, in Sections 3 and 4. The exposition of this
Section 2 was meant to  show that hardly any additional assumption need
to be introduced to reach a quantitative\footnote{Note that $\Lambda^2
(W^2)$ is related \cite{Diff2000,Schi-Ku} to the gluon-structure function. As in 
case of the gluon-structure function, that depends on a more or less arbitrary
input distribution, the three parameters in $\Lambda^2 (W^2)$ cannot 
be theoretically derived at present.} 
description of the experimental data on $\sigma_{\gamma^*p} 
(W^2, Q^2)$, once the $x \to 0$ limit of
the two-gluon-exchange mechanism 
is adopted.

\section{Deeply virtual Compton scattering}
\setcounter{equation}{0}

With the results from Section 2, it is a simple matter to deduce the 
forward-scattering amplitude and the forward-scattering cross section
for DVCS. The reaction $\gamma^*p \to \gamma p$ is described by the
diagrams in fig. 1 upon putting the final photon, $\gamma$, on-shell,
$Q^2 = 0$.

Technically, three different, but equivalent, ways suggest themselves to
derive the (dominant) imaginary part of the DVCS amplitude:
\begin{itemize}
\item[i)] Evaluate the representation (\ref{2.5}) for the transverse 
part\footnote{Helicity conservation is assumed.}
of the total virtual photoabsorption cross section upon substituting
$Q^2 = 0$ in one of the photon wave functions, $\psi_T (r_\bot, z (1-z),
Q^2 = 0)$, in (\ref{2.5}). \footnote{This approach approximates 
the imaginary part of the forward scattering amplitude, i.e. 
the color-dipole cross section at $t_{\rm min}\ne 0$ by its
value at $t=0$, which value is identical to the one that
enters the total virtual photoabsorption
cross section.  The approximation is well justified 
by the exceedingly small value of $|t_{\rm min}|$. 
Compare (\ref{3.16}) below.}
\item[ii)] Use the sum rule (\ref{2.1}) and its GVD interpretation that
associates the factor explicitly shown in (\ref{2.16}) with the propagator of
the $(q \bar q)^{J=1}$ state of mass $M_{q \bar q} \equiv M$ and
four-momentum-squared $q^2= -Q^2$. Simply put $Q^2 = 0$ in this
propagator factor in (\ref{2.1}), and evaluate the integral over
$M^2$.
\item[iii)] Use the symmetry between the incoming and the outgoing photon
in the sum rule (\ref{2.1}) for $\sigma_{\gamma^*_T} (W^2, Q^2)$, and
accordingly put $Q^2 = 0$ in the diffractive forward-scattering amplitude
appearing in (\ref{2.1}) via the square root of the forward-scattering cross
section (\ref{2.3}).
\end{itemize}
Employing method iii)\footnote{For completeness, we also verified our result
by using the technically somewhat more involved methods i) and ii)}, we note
that the diffractive cross section (\ref{2.35}) in the $Q^2 = 0$ limit
reduces to\footnote{The (discontinuous) sudden vanishing of the cross
section (\ref{3.1}) for $M^2 \geq \Lambda^2 (W^2)$ is an artefact of the
$\delta$-function ansatz (\ref{2.24}).} 

\begin{eqnarray}
&& {{d \sigma_{\gamma^*_T p \to (q \bar q)^{J=1}_T p}}\over{dt dM^2}}
\bigg|_{t = 0} (W^2, Q^2 = 0, M^2) = \nonumber \\
& =& {1\over{16 \pi}} 
     {{\alpha R_{e^+e^-}}\over{3 \pi}} (\sigma^{(\infty)})^2 
     \cases{ {1\over{M^2}} &for~$M^2 < \Lambda^2 (W^2)$,\cr 
              0, & for~ $M^2 \geq \Lambda^2 (W^2)$.}
\label{3.1}
\end{eqnarray}
 
Rewriting (\ref{2.1}) as a sum rule for the imaginary part of the forward
scattering amplitude for $\gamma^* p \to \gamma^* p$, we substitute
(\ref{3.1}), and upon squaring the result, we find
\begin{eqnarray}
&& \frac{d \sigma_{\gamma^*p \to \gamma p}}{dt} \bigg|_{t=0} (W^2, Q^2) =
    \nonumber \\
&=&  \frac{1}{16 \pi} \bigg( \frac{\alpha R_{e^+e^-} \sigma^{(\infty)}}{3 \pi}
\bigg)^2 \bigg( \int_{m^2_0}^{\Lambda^2(W^2)} \frac{dM^2}{Q^2 + M^2}
\bigg)^2 (1 + \beta^2_T),
\label{3.2}
\end{eqnarray}
or
\be
\frac{d \sigma_{\gamma^* p \to \gamma p}}{dt} \bigg|_{t=0} (W^2, Q^2)
= \frac{1}{16 \pi} \bigg( \frac{\alpha R_{e^+e^-} \sigma^{(\infty)}}{3 \pi}
\bigg)^2 \bigg( \ln \frac{Q^2 + \Lambda^2 (W^2)}{Q^2 + m^2_0} \bigg)^2
(1 + \beta^2_T).
\label{3.3}
\ee
The factor $1 + \beta^2_T$ in (\ref{3.2}) and (\ref{3.3}) takes care of
the real part of the forward-scattering amplitude. The correction $\beta_T$
for the real part will not be discussed in detail, since it is expected
to be fairly negligible, $\beta^2_T \ll 1$,\footnote{
A recent estimate \cite{favart} in a color-dipole
approach finds a value that decreases from about $\beta_T^2 \lsim 0.2$
for $W\cong 30$ GeV to $\beta_T^2\lsim  0.1$ at $W\cong  300$
GeV.} at least with respect to the
accuracy of the presently available experimental data. A brief comment
concerns the lower limit, $m^2_0$, in (\ref{3.2}). 
As mentioned in connection with the results for the total cross section
and for diffraction, a (correct) symmetric introduction of the threshold
mass $m_0$ with respect to the incoming and outgoing photon leads to 
a correction term \cite{Cvetic}.
 We have checked that this correction only 
insignificantly, at the 1\% level, affects the very simple final result
(\ref{3.3}) for DVCS in the forward direction.

In the low-$Q^2$ limit of $Q^2 \ll \Lambda^2 (W^2)$, we may represent
(\ref{3.3}) in terms of the scaling variable from (\ref{2.31}),
\be
\eta^{-1} (W^2, Q^2) = \frac{\Lambda^2 (W^2)}{Q^2 + m^2_0},
\label{3.4}
\ee
to become
\be
\frac{d \sigma_{\gamma^* p \to \gamma p}}{dt} \bigg|_{t=0}
(W^2, Q^2) = \frac{1}{16 \pi} \bigg( \frac{\alpha R_{e^+e^-} 
\sigma^{(\infty)}}{3 \pi} \bigg)^2 \cdot (\ln \eta^{-1})^2 (1 + \beta_T^2),
\label{3.5}
\ee
or, upon introducing the total virtual photoabsorption cross section from
(\ref{2.29}) on the right-hand side in (\ref{3.5}),
\be
\frac{d \sigma_{\gamma^*p \to \gamma p}}{dt} \bigg|_{t=0} (W^2, Q^2)
= \frac{1}{16 \pi} \sigma^2_{\gamma^*p} (\eta) (1 + \beta^2_T).~~~~~~~~~~~~~~~ (\eta \ll 1)
\label{3.6}
\ee
Even though $Q^2 = 0$ for the outgoing photon in DVCS, for $\Lambda^2 (W^2) 
\gg Q^2$, according
to (\ref{3.6}), we nevertheless have approximate validity of the 
optical theorem; the imaginary part of the forward scattering amplitude
for $\gamma^*p \to \gamma p$ is in good approximation given by the
imaginary part of the amplitude for $\gamma^* p \to \gamma^* p$. In the
limit of real Compton scattering (\ref{3.6}) becomes identical to the
optical theorem for (real) Compton scattering,
\be
\frac{d \sigma_{\gamma p \to \gamma p}}{dt} \bigg|_{t=0} (W^2) =
\frac{1}{16 \pi} \sigma^2_{\gamma p} (W^2) (1 + \beta^2_T).
\label{3.7}
\ee
The transition from real Compton scattering to DVCS for $\eta \ll 1$,
according to (\ref{3.7}) and (\ref{3.6}), corresponds to the 
substitution
\be
\ln \frac{\Lambda^2 (W^2)}{m^2_0} \to \ln 
\frac{\Lambda^2 (W^2)}{Q^2 + m^2_0}.
\label{3.8}
\ee

Turning to the opposite limit of $Q^2 \gg \Lambda^2 (W^2)$, by expanding
the logarithmic function in (\ref{3.3}), we find
\bqa
 \frac{d \sigma_{\gamma^*p \to \gamma p}}{dt} \bigg|_{t=0} (W^2, Q^2) 
   &=& \frac{1}{16 \pi} \bigg( \frac{\alpha R_{e^+e^-} 
       \sigma^{(\infty)}}{3 \pi} \bigg)^2 
       \frac{\Lambda^4 (W^2)}{Q^4} (1 + \beta^2_T)  \nonumber \\
   &=& \frac{1}{16 \pi} \bigg( \frac{\alpha R_{e^+e^-} 
       \sigma^{(\infty)}}{3 \pi} \bigg)^2 \frac{1}{\eta^2} (1 + \beta^2_T).
\label{3.9}
\eqa
The approach to this limit (\ref{3.9}) of a $1/Q^4$ dependence
is slow, i.e. in the HERA   energy range of 30 GeV$\lsim W\lsim$ 300
GeV, the behavior of (\ref{3.3}) is closer to $1/Q^3$ (as found
in the ZEUS experiment\cite{HERA-I}) than to $1/Q^4$.  

In (\ref{3.9}),
again we may introduce the virtual photoabsorption cross section, now
inserting its asymptotic form from (\ref{2.30}),
\bqa
  \frac{d \sigma_{\gamma^*p \to \gamma p}}{dt} \bigg|_{t=0} (W^2, Q^2) 
  &=& \frac{4}{16 \pi} \sigma^2_{\gamma^*p} (W^2, Q^2) (1 + \beta^2_T) 
     \label{3.10} \\
  &=& \frac{9}{16 \pi} \sigma^2_{\gamma^*_T p} (W^2, Q^2) 
     (1 + \beta^2_T), ~~~({\rm for~} Q^2 \gg \Lambda^2 (W^2)).
      \nonumber
\eqa
According to (\ref{3.10}), for asymptotic values of $Q^2 \gg \Lambda ^2 (W^2)$,
the dependence on the kinematical variables of the forward-scattering
amplitude for DVCS, $\gamma^* p \to \gamma p$, is determined by the
amplitude for $\gamma^* p \to \gamma^* p$. The final photon being on shell,
$Q^2 = 0$, in (\ref{3.10}) only leads to a constant, even though significant,
enhancement factor with respect to what is obtained by applying the 
optical theorem
to the amplitude for $\gamma^* p \to \gamma^* p$.

We may combine the limits (\ref{3.6}) and (\ref{3.9}) to conclude
\bq
  \rho_T\equiv
\frac{16 \pi \frac{d \sigma_{\gamma^*p \to \gamma p}}{dt} \bigg|_{t=0}
(W^2, Q^2)}{\sigma^2_{\gamma_T^*p} (W^2, Q^2)} =
\left\{ \begin{array}{r@{\quad,\quad}l}
1 &{\rm for~} Q^2 \ll \Lambda^2 (W^2), \\
9 &{\rm for~} Q^2 \gg \Lambda^2 (W^2).
\end{array} \right. 
\label{3.11}
\eq
If the transverse photoabsorption cross section in (\ref{3.11}) is
replaced by the total one, $\sigma_{\gamma^*} (W^2, Q^2)$, the factor 9
is replaced by 4.
In fig. 3, we show a plot of the ratio
(\ref{3.11}). The plot explicitly displays the enhancement due to putting
the final photon on-shell in the square of the amplitude for $\gamma^* p
\to \gamma^* p$. 
\footnote{The result (\ref{3.11}) and fig.4 are based on the total 
cross section (\ref{2.27}) to (\ref{2.30}) that scales in $\eta$.
Scaling in $\eta$ is violated at large $\eta$ for finite $W$, since the
diffractive mass spectrum has an upper bound, $m_1^2$ \cite{Schi-Ku}.
Taking this effect into account leads to an enhancement of the ratio
(\ref{3.11}) in fig.4.  The enhancement in fig.4 starts at $Q^2\approx
50 GeV^2$ and reaches about 10\% at $Q^2\approx 100 GeV^2$.}
In fig.4, we have indicated the energy, $W=89$ GeV, at which the ratio
in (\ref{3.11}) was evaluated, even though the ratio is 
(obviously) independent of $W$ in the limits indicated in (\ref{3.11}).

The logarithmic behavior (\ref{3.5}) that sets in for $\eta \ll 1$, leads
to the conclusion that
\be
\lim_{W \to \infty \atop Q^2 = {\rm const}} 
\frac{
   {{d \sigma_{\gamma^* p \to \gamma p}}\over{dt}} \bigg|_{t=0} 
   (W^2, Q^2)}   
 {  {{d \sigma_{\gamma p \to \gamma p}}\over{dt}}\bigg|_{t=0} (W^2)} = 1.
\label{3.12}
\ee
The cross section for DVCS at any fixed $Q^2$ for sufficiently high energy,
$W$, approaches the one of real Compton scattering. The result
(\ref{3.12}) is the Compton-scattering analogue of the asymptotic 
relationship (\ref{2.38}) for the total photoabsorption cross section.

We emphasize that ``saturation'' in the sense of 
(\ref{2.38}) and (\ref{3.12}) does not
depend on a specific ``saturation-model'' assumption. Saturation in the
sense of (\ref{2.38})) and (\ref{3.12}) only rests on the underlying 
two-gluon exchange
generic structure\footnote{If anything in addition, it is the convergence of
the integrals (\ref{2.22}) and (\ref{2.23}) over the distribution in the
gluon transverse momentum and its first moment that enters (\ref{2.38}) 
and (\ref{3.12}).} from fig. 1 that implies the basic relations 
(\ref{2.21}) to (\ref{2.23}).

The experiments\cite{HERA-I} on $\gamma^* p \to \gamma p$ do not extract
so far the forward-production cross section, nor the slope in momentum
transfer. Theoretical considerations on the slope are beyond the scope of
the present work. Assuming an exponential fall-off with slope $b$,
${\rm exp} (bt)$, we obtain from (\ref{3.3})
\be
\sigma_{\gamma^* p \to \gamma p} (W^2, Q^2) = \frac{1}{16 \pi b}
\left( \frac{\alpha R_{e^+e^-} \sigma^{(\infty)}}{3 \pi} \right)^2
\left(\ln \frac{Q^2 + \Lambda^2 (W^2)}{Q^2 + m^2_0} \right)^2 (1 + \beta^2_T).
\label{3.13}
\ee

In the comparison with the experimental data from HERA, 
we proceed in two steps.  In the first step, we put 
\be
\beta_T = 0,~~~b = 4~ {\rm GeV}^{-2}.
\label{3.14}
\ee
All other quantities in (\ref{3.13}), the product $R_{e^+e^-} 
\sigma^{(\infty)}$, as well as $\Lambda^2 (W^2)$ and $m^2_0$, were fixed
by the analysis of the total cross section, $\sigma_{\gamma^*p} (W^2, Q^2)$.
Compare (\ref{2.32}) to (\ref{2.34}).

When looking at the comparison with the experimental data
in figs.4 and 5, the drastic assumption of a constant slope 
$b$ in (\ref{3.14}) has to be kept in mind, and the data 
indeed indicate
\footnote{This was previously observed by the ZEUS collaboration 
and in the theoretical analysis in \cite{FMS}.}
that the $Q^2$ dependence requires a decrease in $b$ with 
increasing $Q^2$, as expected from the effective decrease of 
the dipole size $r_\perp$ with increasing $Q^2$ contained in the
photon lightcone wave function (for the initial photon) 
in (\ref{2.5}).
 
  In order to quantitatively estimate the influence of 
a change in slope with $Q^2$, we assume a dependence of
$b(Q^2)$ on $Q^2$ extracted from $\rho^0$ electroproduction, 
even though the mass spectrum of the intermediate state coupled to 
the outgoing real photon extends quite far beyond 
the $\rho^0$ mass.\footnote{An estimate of the $\rho^0$ 
contribution to DVCS gave a contribution of the order of 
20\% \cite{dgs} at low $Q^2$, rapidly decreasing with increasing $Q^2$.}
The fit to $\rho^0$ electroproduction gave\cite{kreisel}
(in units of GeV$^{-2}$)
\bq
     b_{\rho_0}(Q^2) = {{3.46}\over{M_{\rho^0}^2\Bigl(
                1+ {{\bigl(Q^2/M_{\rho^0}^2\bigr)^{0.74} }\over{2.16}}
              \Bigr) }}+4.21.   \label{3.15}
\eq
Numerically, according to (\ref{3.15}), the slope $b$
decreases from around 10 GeV$^{-2}$ at $Q^2=0$ to about
4.5 GeV$^{-2}$ at $Q^2\cong 100$ GeV$^2$.  
The effect of this change in slope is shown by the dotted line 
in fig.5.

\vspace{0.2 cm}

In conjunction with the difference in slope between (\ref{3.14})
and (\ref{3.15}), the absolute normalization of the cross
section was adjusted by including a real part of $\beta_T^2=0.1$ 
and by an increase of $\sigma^{(\infty)}$ by about 16\%.


Altogether, the validity of the simple formula (\ref{3.3}), 
without any adjusted parameter, is satisfactory and supports
the underlying two-gluon-exchange structure from QCD. As mentioned, the
ansatz (\ref{2.24}) is a technical one that can be adopted without loss
of generality. The specific form of $\Lambda^2 (W^2)$ in (\ref{2.32}) and
(\ref{2.33}) in the appropriate kinematic domain of $\eta \gg 1$ corresponds
\cite{Schi-Ku} to a simple assumption on the gluon distribution. 
An analogous assumption is
inherently contained in any approach to DIS, usually in terms of an
arbitrary input for the gluon distribution.

We end this section with a few comments on other approaches to DVCS. 

Various approaches to DVCS, \cite{DD, FKS} and \cite{MFGS} start from
the color-dipole formulation (\ref{2.11}) for the total cross section
that does not constrain the restriction in the color-dipole cross section
to $(q \bar q)^{J=1} p \to (q \bar q)^{J=1} p$ transitions. 
The observed differences \cite{Landshoff} between dipole cross sections
leading to equivalent predictions for $\gamma^* p \to \gamma^* p$ and
$\gamma^* p \to \gamma p$ are presumably due to $J \not= 1$ contributions.
Both approaches of refs. \cite{DD} and \cite{FKS} use the energy,
$W$, as the basic variable the dipole-cross section depends on, while
ref. \cite{MFGS} employs a dependence on the two variables $x$ and $Q^2$.

Both refs. \cite{DD} and \cite{FKS} use a two-component approach (soft
and hard Pomeron) in the ansatz for the dipole cross section to accomodate
both the $Q^2 \to 0$ and the $Q^2 \to \infty$ limits in distinction from
our one-component approach (\ref{2.25}) that describes the hard and the
soft limit by the transition from $1/\eta$ to $\ln (1/\eta)$ according to
(\ref{2.29}) and (\ref{2.30}).  Our formulation is unique with respect to
the simplicity of the final result for DVCS in (\ref{3.3}) including the
$Q^2 \ll \Lambda^2 (W^2)$ and $Q^2 \gg \Lambda^2 (W^2)$ limits in 
(\ref{3.5}) and (\ref{3.9}), and with respect to
the transparent connection with the
total photoabsorption cross section in (\ref{3.6}) and (\ref{3.10}).

     The approach of the present paper and related work 
on DVCS based on the color-dipole picture 
\cite{DD} neglects the effect of a 
non-zero minimal momentum transfer to the proton,
$t_{\rm min}$,
resulting from the difference in four-momentum
squared of the incoming ($Q^2>0$) and outgoing photon
($Q^2=0$) in (forward) DVCS.  For $\gamma^*p \to Xp$,
in leading order in $(Q^2+M_X^2)/W^2$, 
\be
t_{min} \simeq - \frac{(Q^2 + M^2_X)^2}{W^4} M^2_p,
\label{3.16}
\ee
where $M_p$ denotes the proton mass.
For DVCS, where $M^2_X = 0$, we have
$|t_{min}| = x^2 M^2_p$, and, accordingly, $|t_{min}|$ 
is of negligible magnitude with respect to all other dimensionful
variables, since it lies between $10^{-2} GeV^2$ at $x = 10^{-1}$ and
$10^{-8} GeV^2$ at $x = 10^{-4}$.
The effect  of $P^\prime \ne P$, where 
$P$ and $P^\prime$ refer to the
incoming and outgoing nucleon, respectively,
has been investigated by generalizing parton 
distributions and evolution equations to this case of 
$P^\prime \ne P$\cite{gpd,collins}. 
Generalized parton distributions (GPD's) can thus be
predicted at any $Q^2$, once they have been specified
at  an input scale, $Q_0^2$, by an appropriate
parameterization.  For DVCS see \cite{FFS,dgs,FMS}.
A comparison of the 
theoretical prediction thus obtained \cite{FMS, HERA-I} 
for the $Q^2$ 
dependence of DVCS  with ours in fig.5 reveals no drastic 
difference.
A theoretical analysis of (forward) DVCS that 
only relies on those parameters that are
fixed by measurements of 
$\sigma_{\gamma^*p}(W^2,Q^2)$, thus ignoring the change in
four-momentum of the proton, $P \ne P^\prime$, so far
seems adequate for a representation
of the available experimental data on DVCS.

\section{Vector-Meson Electroproduction}
\setcounter{equation}{0}

Using the results for diffractive production, $\gamma^* p \to (q \bar q)^{J=1}
p$, of the $(q \bar q)^{J=1}$ continuum collected in (\ref{2.35}) to
(\ref{2.37}) in Section 2, and applying quark-hadron duality, it is now
a simple matter to arrive at a parameter-free prediction for vector-meson
forward production.

Quark-hadron duality \cite{Sakurai} says that the asymptotic cross section
for $e^+e^-$ annihilation into a quark-antiquark pair, $e^+ e^- \to q \bar q$,
interpolates the production of the low-lying vector mesons of the corresponding
quark flavor. That is, the integrals over the low-lying vector meson peaks in
$e^+e^-$ annihilation, when averaged over an interval 
$\Delta M^2_V$ determined
by the vector-meson level spacing, become identical to the low-energy
continuation of the asymptotic cross sections for $e^+e^- \to q \bar q$ with
appropriate flavor of the quark $q$.

For the case of diffractive production, quark-hadron duality allows us to
determine the cross section for vector-meson production by integration of
the cross sections (\ref{2.35}) to (\ref{2.37}) for the diffractively
produced $(q \bar q)^{J=1}$ continuum\footnote{Compare also \cite{Levin} for
the application of quark-hadron duality to vector-meson production.}\footnote{
A refined treatment should simultaneously analyse quark-hadron duality in 
$e^+ e^-$ annihilation and diffractive production},
\be
\frac{d \sigma_{\gamma^*p \to (q \bar q)^{J=1}_{M(V)}p}}{dt} \bigg|_{t = 0}
(W^2, Q^2) = \int dz\int_{\Delta M^2_V} dM^2
\frac{d \sigma_{\gamma^*p \to 
(q \bar q)^{J=1} p}}{dt dM^2dz} \bigg|_{t=0} (W^2, Q^2, M^2).
\label{4.1}
\ee
In (\ref{4.1}), $\gamma^*p \to (q \bar q)^{J=1}_{M(V)}p$, may refer to the
sum of transverse and longitudinal cross sections, as well as to transverse
and longitudinal cross sections separately, whereby $\gamma^*$ has to
be replaced by $\gamma^*_T$ and $\gamma^*_L$, respectively\footnote{We
restrict ourselves to evaluating $\gamma^*_T p \to (q \bar q)^{J=1}_T p$ and 
$\gamma^*_L p \to (q \bar q)^{J=1}_L p$, i.e. by disregarding transitions
such as $\gamma^*_T p \to (q \bar q)^{J=1}_L p$, we assume helicity
conservation.}. The mass of the vector meson $V$ is denoted by $M(V) \equiv
M_V$, and the quark flavor of the cross section
on the right hand side in (\ref{4.1}) has to be identical to
the quark content of the vector meson, i.e. when substituting 
(\ref{2.35}) to (\ref{2.37}) into (\ref{4.1}), $R_{e^+e^-}$ is to be 
replaced by 
\bq
R_{e^+e^-} = 3 \sum_q Q^2_q \to
\left\{ \begin{array}{r@{\quad=\quad}l}
R^{(\rho^0)} = \frac{9}{10} R^{(\rho^0 + \omega)} & \frac{3}{2}, \\
R^{(\omega)} = \frac{1}{10} R^{(\rho^0 + \omega)} & \frac{1}{6},\\
R^{(\phi)} = R^{(\Upsilon)} & \frac{1}{3},\\
R^{(J/\psi)} & \frac{4}{3}.
        \end{array} \right. 
\label{4.2}
\eq

   When predicting vector-meson production according to
(\ref{4.1}), we have to discriminate between the low-lying 
vector-mesons, $\rho,\omega,\phi$ that may be described by the
limit of vanishing quark mass, $m_u=m_d=m_s=0$, and the vector
mesons, $J/\psi$ and $\Upsilon$, where the approximations 
$M_{J/\psi}^2\approx 4m_c^2$ and $M_\Upsilon^2\approx 4m_b^2$ are 
appropriate.  In Section 4.1 we treat  the light vector mesons by
inserting (\ref{2.35}) to (\ref{2.37}) into (\ref{4.1}).
For the treatment of the heavy vector mesons in Section 4.2, 
we shall generalize (\ref{2.35}) to (\ref{2.37}) to the case
of massive quarks.  A comparison with the
experimental data will be given in Section 4.3.

\subsection{Massless~quarks,~$\rho^0$,$\omega$, $\phi$~ production}

Before giving the explicit result obtained by integration of (\ref{4.1}), it
will be illuminating to consider the approximation of (\ref{4.1}) 
\be
\frac{d \sigma_{\gamma^* p \to (q \bar q)^{J=1}_{M(V)} p}}{dt}\bigg|_{t=0}
(W^2, Q^2) \simeq \Delta M^2_V \int dz \frac{d 
\sigma_{\gamma^*p \to (q \bar q)^{J=1}p}}{dt dM^2dz} \bigg|_{t=0} 
(W^2, Q^2, M^2 = M^2_V),
\label{4.3}
\ee
and determine its behavior for $Q^2 \to \infty$ and $Q^2 \to 0$ upon inserting 
the cross sections from (\ref{2.35}) to (\ref{2.37}) at $M^2=M_V^2$.

The cross sections (\ref{2.35}) to (\ref{2.37}), with respect to $Q^2$, 
dominantly depend on the variable $Q^2 + M^2$. We first of all consider the
limit of
\be
Q^2 + M^2_V \gg \Lambda^2 (W^2).
\label{4.4}
\ee
For definiteness, we note that according to (\ref{2.32})
\be
2 GeV^2 \lsim \Lambda^2 (W^2) \lsim 7 GeV^2
\label{4.5}
\ee
for the energy range covered by HERA of $30 GeV \lsim W \lsim 300 GeV$.
From the leading term in the expansion of (\ref{2.37}) in powers of 
$\Lambda^2 (W^2)/(Q^2 + M^2_V)$, upon insertion in (\ref{4.3}) we find
\bqa
  &&  \frac{d \sigma_{\gamma^* p \to (q \bar q)^{J=1}_{M(V)} p}}{dt} 
    \bigg|_{t=0} (W^2, Q^2, M^2_V)  \nonumber \\
  & = & \frac{1}{16 \pi} \frac{\alpha R^{(V)}}{3 \pi}
    (\sigma^{(\infty)}(W^2))^2 
    \frac{\Lambda^4 (W^2) \Delta M^2_V}{(Q^2 + M^2_V)^3}
   {{Q^2}\over{Q^2+M_V^2}}.
\label{4.6}
\eqa
The transverse production cross section, from the expansion of (\ref{2.35}),
is given by
\bqa
  {{d \sigma_{\gamma^*_T p \to (q \bar q)^{J=1}_{M(V)T}p}}\over{dt}} 
        \bigg|_{t=0} (W^2, Q^2, M^2_V) 
   & = & \frac{1}{16 \pi} \frac{\alpha R^{(V)}}{3 \pi}
    (\sigma^{(\infty)}(W^2))^2  \label{4.7} \\
     & &\cdot \frac {4 \Lambda^4 (W^2) M^2_V \Delta M^2_V} {(Q^2
+ M^2_V)^4} \Bigl({{Q^2}\over{Q^2+M_V^2}}\Bigr)^2.
\nonumber 
\eqa

The results (\ref{4.6}) and (\ref{4.7}), according to (\ref{4.4}), are 
not only valid for the production of light vector mesons, but 
their range of validity also
includes the case of the continuum of heavy $q \bar q$ states formed
from light quarks with
\bq
     M_V^2 \equiv M_{q\bar q}^2 \gg \Lambda^2(W^2),  \label{4.8}
\eq
where (\ref{4.4}) is valid even for $Q^2\ge 0$.

We turn to $\rho^0(\omega,\phi)$ production, where (\ref{4.4})
with (\ref{4.5}) requires 
\bq
     Q^2 \gg \Lambda^2(W^2)  \label{4.9}
\eq
since $\Lambda^2(W^2) > M_{\rho^0(\omega,\phi)}^2$.
The unpolarized cross section (\ref{4.6}) in this case 
may be more appropriately
written as
\bqa
 &&\frac{d \sigma_{\gamma^*p \to (q \bar q)^{J=1}_{M(V)} p}}{dt} \bigg|_{t=0}
(W^2, Q^2, M^2_V) = \frac{1}{16 \pi} \frac{\alpha R^{(V)}}{3 \pi}
(\sigma^{(\infty)}(W^2))^2 
\frac{\Lambda^4 (W^2) \Delta M^2_V}{(Q^2 + M^2_V)^3}.
\nonumber \\
   & & ~~~~~~~~~~~~~~~~~~~~\bigl(V=(\rho^0, \omega,\phi); Q^2\gg
      \Lambda^2(W^2)>M_{\rho^0,\omega,\phi}^2\bigr) \label{4.10}
\eqa

Since the transverse cross section (\ref{4.7}) is suppressed by a
power of $Q^2+M_V^2$, the longitudinal cross section in its leading
term coincides with (\ref{4.6}) and the longitudinal-to-transverse
ratio in the asymptotic limit (\ref{4.4}) is given by
\bq
     R_{L/T} = {{Q^2}\over{4M^2_V}}, 
     ~~~~~ (Q^2\gg\Lambda^2(W^2)>M_V^2).
\label{4.11}
\eq

For the opposite limit of 
\bq
  Q^2+M_V^2\ll \Lambda^2(W^2),\label{4.12}
\eq
by expansion of (\ref{2.37}), and integration from $M^2_1$ to 
$M^2_1 + \Delta M^2_V$, we find 
\be
\frac{d \sigma_{\gamma^* p \to (q \bar q)^{J=1}p}}{dt} \bigg|_{t=0}
(W^2, Q^2, M^2_V) = \frac{1}{16 \pi} \frac{\alpha R^{(V)}}{3 \pi} 
(\sigma^{(\infty)}(W^2))^2 \ln 
\left( 1 + \frac{\Delta M^2_V}{Q^2 + M^2_1}\right).
\label{4.13}
\ee

\vspace{0.2 cm}

A final remark concerns the energy dependence of the whole  diffractively
produced  $(q \bar q)^{J=1}$ continuum in the limit (\ref{4.12}),
relevant for photoproduction of light quarks, $q = u,d,s$ at HERA
energies. With
(\ref{3.1}), we find,
\bqa
   & &\int_{m^2_0} dM^2 
     \frac{d \sigma_{\gamma^*p \to (q \bar q)^{J=1}_T p}}{dt dM^2} 
     \bigg|_{t=0} (W^2, Q^2 = 0,M^2) \nonumber \\
   &=& {1\over{16 \pi}} {{\alpha R_{e^+e^-}}\over{3 \pi}} 
       \left( \sigma^{(\infty)} (W^2)\right)^2 
       \int^{\Lambda^2 (W^2)}_{m^2_0} \frac{dM^2}{M^2}  \nonumber \\
   &=& {1\over{16 \pi}} {{\alpha R_{e^+e^-}}\over{3 \pi}} 
       \left( \sigma^{(\infty)} (W^2) \right)^2 
       \ln \frac{\Lambda^2(W^2)}{m^2_0}.  \label{4.14}
\eqa
While the energy dependence in photoproduction of a discrete vector-meson 
state (\ref{4.13}) is even 
weaker than the energy dependence of the total photoproduction cross section,
the energy dependence of the whole $(q \bar q)^{J=1}$ continuum, 
with the approximate constancy
of $\sigma^{(\infty)} (W^2)$, 
shows the logarithmic
dependence of $\sigma_{\gamma p} (W^2)$ from (\ref{2.29}).

We finally give the expressions for the vector-meson-production
cross section obtained by evaluating the quark-hadron-duality relation
 (\ref{4.1})
without further approximation. 
Upon insertion of
(\ref{2.35}) to (\ref{2.37}), and integration over $dz$ and $dM^2$, we find
\be
\frac{d \sigma_{\gamma^*p \to (q \bar q)^{J=1}_{M(V)}p}}{dt} \bigg|_{t=0}
(W^2, Q^2) = \frac{1}{16 \pi} \frac{\alpha R^{(V)}}{3 \pi}
\left( \sigma^{(\infty)} \right)^2 \left[ \Pi (\Lambda^2(W^2), Q^2, M^2)
\right]^{M^2_2}_{M^2_1}
\label{4.15}
\ee
where the function $\Pi (\Lambda^2 (W^2), Q^2, M^2)$ is to be evaluated
at the limits of $M^2_1$ and $M^2_2$, where
\begin{eqnarray}
& \Delta M^2_V = M^2_2 - M^2_1, \nonumber \\
&M^2_1 < M^2_V < M^2_2.
\label{4.16}
\end{eqnarray}
For the case of the 
sum of transverse and longitudinal production cross sections, we have
{\footnotesize
\begin{eqnarray}
  \Pi (\Lambda^2 (W^2), Q^2, M^2) &=& + {1\over 2} \ln 
{{(\Lambda^2+Q^2)(\sqrt X+Q^2+\Lambda^2)
             +M^2(Q^2-\Lambda^2)}\over
            {\sqrt X + M^2+Q^2-\Lambda^2}} \label{4.17} \\
        & &-{{\Lambda^2}\over{\sqrt{\Lambda^2(4Q^2+\Lambda^2)}}}
           \ln {{\sqrt{\Lambda^2(4Q^2+\Lambda^2)}\sqrt X
             +\Lambda^2(3Q^2-M^2+\Lambda^2)} \over{M^2+Q^2}},
           \nonumber
\end{eqnarray} }
while for the longitudinal and the transverse case, separately,
{\footnotesize
\begin{eqnarray}
  \Pi_L (\Lambda^2 (W^2), Q^2, M^2) &=&- {1\over{M^2+Q^2}}
           +{1\over {2\sqrt{Q^2\Lambda^2}}}\arctan
           {{M^2+Q^2-\Lambda^2}\over{2\sqrt{Q^2\Lambda^2}}} 
             \label{4.18} \\
        & &+{2\over{\sqrt{\Lambda^2(4Q^2+\Lambda^2)}}}
           \ln {{\sqrt{\Lambda^2(4Q^2+\Lambda^2)}\sqrt X
             +\Lambda^2(3Q^2-M^2+\Lambda^2)} \over{M^2+Q^2}},
             \nonumber 
\end{eqnarray}}
where $\Lambda^2$ stands for $\Lambda^2(W^2)$ and $X$ is given by
\be
X \equiv  (Q^2+M^2-\Lambda^2)^2+4Q^2\Lambda^2,
\label{4.19}
\ee
and
{\footnotesize
\begin{eqnarray}
 \Pi_T (\Lambda^2 (W^2), Q^2, M^2) &=& {{Q^2}\over{M^2+Q^2}}
           -{1\over 2}\sqrt{{Q^2}\over{\Lambda^2}}\arctan
           {{M^2+Q^2-\Lambda^2}\over{2\sqrt{Q^2\Lambda^2}}} 
             \label{4.20} \\
        & &+ {1\over 2} \ln {{(\Lambda^2+Q^2)(\sqrt X+Q^2+\Lambda^2)
             +M^2(Q^2-\Lambda^2)}\over
            {\sqrt X + M^2+Q^2-\Lambda^2}} \nonumber \\
        & &-{{2Q^2+\Lambda^2}\over{\sqrt{\Lambda^2(4Q^2+\Lambda^2)}}}
           \ln {{\sqrt{\Lambda^2(4Q^2+\Lambda^2)}\sqrt X
             +\Lambda^2(3Q^2-M^2+\Lambda^2)} \over{M^2+Q^2}}.
             \nonumber 
\end{eqnarray}}

Upon adopting an exponential $t$-dependence,  $\exp (b_Vt)$, the cross
section for vector meson production, $\sigma_{\gamma*{p \to 
(q \bar q)^{J=1}_{M(V)} p} (W^2, Q^2)}$, is obtained from (\ref{4.3}) and
(\ref{4.15}) by
multiplication by $1/b_V$.

For the case of the $\rho^0$ meson we have compared the results of the 
integration of (\ref{4.1}) given by (\ref{4.15}) with the approximate results 
for large $Q^2$ in (\ref{4.10}) and for small $Q^2$ in (\ref{4.13}). 
With $M^2_{\rho^0} = 0.59\,{\rm GeV}^2$ and $\Delta M^2_{\rho^2} = 
1 \, {\rm GeV}^2$ and the 
lower bound $M^2_1 = 0.36 \, {\rm GeV}^2$ in (\ref{4.15}), we find that the 
large-$Q^2$ approximation for $Q^2 \ge 90 \, {\rm GeV}^2$ in (\ref{4.10})
overestimates the exact evaluation of (\ref{4.1}) by less than 10\%. 
With decreasing $Q^2$ the large-$Q^2$ approximation substantially 
overestimates the exact evaluation, since (\ref{4.4}) becomes violated and 
in addition the 
approximation of $Q^2 + M^2$ by a constant, $Q^2 + M^2_V$, becomes less
justified. For $Q^2 \rightarrow 0$, the approximation (\ref{4.13}) 
exceeds the exact evaluation by a few percent.
Altogether, for semi-quantitive discussions, the approximations (\ref{4.10})
and (\ref{4.13}) are very useful, while for detailed comparison with the 
experimental data the results based on (\ref{4.15}) should be employed.

\subsection{ Massive~quarks, $J/\psi$ and $\Upsilon$~production}
The light-cone wave functions (\ref{2.8}) and (\ref{2.9}) 
include a non-zero rest mass, 
$m_q$, of the quark. 
There are essentially two important effects, when passing from the 
approximation of massless quarks, relevant for $\rho^0 , w , \phi$ 
production, where $M^2_V \gg 4 m^2_q \cong 0$, to massive quarks, relevant
for $J/\psi$ and $\Upsilon$ production, where $M^2_V \cong 4m^2_q \not= 0$.

First of all, the transition from massless to massive quarks affects the 
lightcone variable $z$. In the massless-quark case, we have 
\be
0 \le z \le 1 , 
\label{4.21}
\ee
and $z$ is related to the angle of the $q \bar q$ axis relative to the 
photon-direction in the $q \bar q$ rest frame via 
\begin{eqnarray}
\sin\theta & = & 2 \sqrt{z(1-z)} , \nonumber \\
\cos\theta & = & 1 - 2 z . 
\label{4.22}
\end{eqnarray}
In the case of massive quarks, (\ref{2.18}), the range of $z$ 
for given quark mass and given mass $M_{q \bar q}$ is determined by 
\be
M^2 \equiv M^2_{q \bar q} = \frac{m^2_q}{z(1-z)}.  
\label{4.23}
\ee 
One finds
\bqa
    &&   z_-\le z \le z_+, \nonumber \\
    &&  z_\pm = \frac{1}{2} \pm \frac{1}{2} 
       \sqrt{1 - 4 \frac{m^2_q}{M^2_{q \bar q}}}, 
\label{4.24}
\eqa
and an integration over $dz$ is restricted to the interval 
\be
\Delta z = z_+-z_- =\sqrt{1-4 \frac{m^2_q}{M^2_{q \bar q}}} = \left\{ 
\matrix{1 & {\rm for} & m^2_q = 0 , \cr
0, & {\rm for} & m^2_q = \frac{1}{4} M^2_{q \bar q} . } \right.
\label{4.25}
\ee
Upon introducing the variable $y$ defined by 
\be
z = (z_+ - z_-) y + z_-
= \left\{ 
\matrix{z_+ , & {\rm for} & y = 1, \cr
z_- , & {\rm for} & y = 0 , } \right.
\label{4.26}
\ee
the integration over $dz$ in the massive-quark case is represented by 
\be
\int^{z_+}_{z_-} \, dz = \sqrt{1 - \frac{4m^2_q}{M^2_{q \bar q}}} \int^1_0 \,
dy .
\label{4.27}
\ee
For later reference, we note the integral over the threshold factor in 
(\ref{4.27}), 
\begin{eqnarray}
 \Delta F^2(m_q^2,\Delta M_V^2)&\equiv & \int^{4m^2_q + \Delta M^2_V}_{4m^2_q} \, 
     d M^2 \sqrt{1-\frac{4m^2_q}{M^2}} \int^1_0 \, dy  \label{4.28} \\
  &=& (4m^2_q + \Delta M^2_V) \sqrt{\frac{\Delta M^2_V}
     {4m^2_q + \Delta M^2_V}} + 2 m^2_q \, 
      \ln \frac{1 - \sqrt{\frac{\Delta M^2_V}{4m^2_q + \Delta 
      M^2_V}}}{1 + \sqrt{\frac{\Delta M^2_V}{4m^2_q + \Delta M^2_V}}}  .
\nonumber
\end{eqnarray}

In passing, we mention that the $q \bar q$ rest-frame angle in the massive case is
related to $y$ by 
\begin{eqnarray}
\sin\theta & = & 2 \sqrt{y(1-y)} , \nonumber \\
\cos\theta & = & 1 - 2y . 
\label{4.29}
\end{eqnarray}
Moreover, in terms of $y$,
\begin{eqnarray}
z (1-z) & = & \frac{m^2_q}{M^2} + \left( 1 - \frac{4 m^2_q}{M^2} \right) y (1-y),
\nonumber \\
z^2 + (1 - z)^2 & = & \frac{2 m^2_q}{M^2} + \left( 1 - \frac{4 m^2_q}{M^2} 
\right) (y^2 + (1 - y)^2 ) ,
\label{4.30}
\end{eqnarray}
such that $y$ is indeed the appropriate generalization of the variable 
$z$ to the massive-quark case.

The second important modification with respect to the massless-quark case 
concerns the diffractive production cross sections
(\ref{2.35}) to (\ref{2.37})
when passing from 
$M^2_V \gg 4 m^2_q \cong 0$ to $M^2_V \cong 4 m^2_q \not= 0$.
For $q \bar q$ production near threshold this modification
may be accomplished by a simple substitution  to be applied 
in the sum of the transverse and the longitudinal cross section for the 
massless-quark case given in (\ref{2.35}) and (\ref{2.36}).

To derive the substitution prescription, we consider the sum of the 
transverse and 
the longitudinal lightcone wave function given in (\ref{2.8}) and 
(\ref{2.9}), 
respectively, and compare the massive case, $m^2_q \not= 0$, with the 
massless one, $m^2_q = 0$. We find that at production threshold, where
\be
M^2_{q \bar q} = 4m^2_q , \,\,\,\, z(1-z) = \frac{1}{4} , 
\label{4.31}
\ee
the sum of the transverse and longitudinal wave functions of the massive 
case is recovered from the massless one by carrying out the substitution
\be
Q^2 \rightarrow Q^2 + 4 m^2_q ,
\label{4.32}
\ee
in the expression for the massless case. For a vector meson at or closely above 
threshold, the quark mass according to (\ref{4.32}) acts as an 
additive contribution to $Q^2$ of magnitude $4m^2_q \cong M^2_V$.
When applying the substitution (\ref{4.32}) to the sum of the 
diffractive-production differential cross sections for the massless-quark case in 
(\ref{2.35}) and (\ref{2.36}) at $z(1-z)={1\over 4}$, the mass 
$M^2 \equiv \vec k^{~2}_\bot / z (1-z)$ in (\ref{2.35}) and (\ref{2.36})
must be put to zero, in order to correctly realize the threshold 
relation (\ref{4.31}). Finally upon substitution, 
the threshold mass, 
$4 m^2_q$, is to be replaced by the  vector-meson mass, i.e. 
\bqa
(4 m^2_c , 4 m^2_b) \rightarrow (M^2_{J/\psi} , M^2_\Upsilon ). \label{4.33}
\eqa

In order to obtain an approximate expression for the production cross 
section in the case of $M^2_V \cong 4 m^2_q$ when applying quark-hadron
duality according to (\ref{4.1}), we will 
use (\ref{4.27}) and (\ref{4.28}), 
while approximating the $(q \bar q)^{J=1}$ 
production-cross sections (\ref{2.35}) and (\ref{2.36}) upon 
substitution of (\ref{4.32}) by their values 
at threshold, $4 m^2_q$, whereby 
identifying threshold and vector-meson mass, $4m^2_q = M^2_V$.

Upon having carried out the preceeding steps in the sum of the cross 
sections (\ref{2.35}) and (\ref{2.36}), and  upon substituting 
(\ref{4.28}), we find that (\ref{4.1}) is approximated by
\bqa
\frac{d\sigma_{\gamma^* p\rightarrow Vp}}{dt} \Bigg|_{t=0} & = & 
\frac{d\sigma_{\gamma^*_T p \rightarrow V_T p}}{dt} \Bigg|_{t=0} +
\frac{d\sigma_{\gamma^*_L p \rightarrow V_L p}}{dt} \Bigg|_{t=0} 
   \nonumber \\
& = & \frac{3}{2} \cdot \frac{\alpha R^{(V)}}{3 \cdot 16 \pi^2} 
(\sigma^{(\infty)})^2 
\int^{4m^2_q + \Delta M^2_V}_{4m^2_q} \, dM^2 \sqrt{1-\frac{4m^2_q}{M^2}} 
~(Q^2 + M^2_V) \nonumber \\
   & \cdot & \left( \frac{1}{Q^2 + M^2_V } 
      - \frac{1}{Q^2 + M^2_V + \Lambda^2 (W^2)} \right)^2, \label{4.34}
\eqa
where $V = J/\psi , Y$. 
The integration over $dM^2$ in (\ref{4.34}) runs from the threshold, $4m^2_q$, 
to the upper limit $4m^2_q + \Delta M^2_V$, where $\Delta M^2_V$ is 
determined by the vector-meson level spacing and $4m^2_q \le M^2_V \le 
4m^2_q + M^2_V$.

The integrand in (\ref{4.34}) is easily verified to be identical to what
one obtains by applying the substitution (\ref{4.32}) 
only to the longitudinal
production cross section (\ref{2.36}). 
In fact, one finds that the transverse cross 
section (\ref{2.35}) vanishes upon applying the substitution
procedure based on 
(\ref{4.32}). 

Identifying the longitudinal cross section in (\ref{4.34})
via its $Q^2$ dependence, (\ref{4.34}) may be rewritten as
\bq
   \frac{d\sigma_{\gamma^*_T p \rightarrow V_T p}}{dt} \Bigg|_{t=0} +  
   \frac{d\sigma_{\gamma^*_L p \rightarrow V_L p}}{dt} \Bigg|_{t=0} 
     = 
   \frac{d\sigma_{\gamma^*_L p \rightarrow V_L p}}{dt} \Bigg|_{t=0}  
   \left( 1 + \frac{M^2_V}{Q^2} \right) ,  \label{4.35}    
\eq
i.e. the longitudinal-to-transverse ratio fulfills 
\be
R_{L/T} = \frac{Q^2}{M^2_V} , \,\,\, ({\rm for}~ Q^2 \ge 0) .
\label{4.36}  
\ee
This ratio differs from the one in the massless-quark case (\ref{4.11}) by the 
missing factor $1/4$ and by the range of validity in $Q^2$.

The final expression for the production cross section (\ref{4.34}) becomes 
\begin{eqnarray}
\frac{d\sigma_{\gamma^* p\rightarrow Vp}}{dt} \Bigg|_{t=0} & = & 
\frac{3}{2}\cdot \frac{\alpha R^{(V)}}{3 \cdot 16 \pi^2} (\sigma^{(\infty)})^2
\frac{\Lambda^4 (W^2)}{(Q^2 + M^2_V)^3} 
\frac{1}{\left( 1 + \frac{\Lambda^2 (W^2)}
{Q^2 + M^2_V}\right)^2} \cdot \nonumber \\
&\cdot & \Delta F^2 (m^2_q , \Delta M^2_V)~~~~~~~~~ (V = J / \psi , \Upsilon)
, \label{4.37}  
\end{eqnarray}
where $\Delta F^2 (m^2_q , \Delta M^2_V)$ is given by (\ref{4.28}).
Note that this expression for the cross section is independent of 
an assumption on 
the relative magnitude of $Q^2 + M^2_V$ and $\Lambda^2 (W^2)$, in 
distinction from the massless-quark case, where (\ref{4.13}) as well as 
(\ref{4.6}) and (\ref{4.10}) are relevant, 
respectively, for $Q^2 + M^2_V \ll \Lambda^2 (W^2)$ and 
$Q^2 + M^2_V \gg \Lambda^2 (W^2)$.
The additional factor 3/2 in the massive-quark case relative to the massless
one is a genuine consequence of the fact that $4m^2_q \cong M^2_V$ for the
heavy vector mesons. While according to (\ref{4.31}), for massive quarks
$z(1-z) = 1/4$, for massless quarks the integration over $z(1-z)$ yield 1/6.
In addition to the factor 3/2 and the factor $R^V$, for 
$Q^2 + M^2_V \gg \Lambda^2 (W^2)$, the massless and the massive case
differ by the replacement of $\Delta M^2_V$ by 
$\Delta F^2(m_q^2, \Delta M_V^2)$.

We point out a few outstanding features of the 
(approximate) result (\ref{4.37}) and its 
massless-quark counterparts (\ref{4.10}) and (\ref{4.13}). 
Rather than referring to  the 
massless-quark and the massive-quark case, we will refer to the 
$\rho^0$ case and the $J/\psi$ case, where $\rho^0$ stands for
$\rho^0$, $\omega$, $\phi$ and $J/\psi$ stands for $J/\psi$ and 
$\Upsilon$. From 
(\ref{4.37}), as well as (\ref{4.10}) and (\ref{4.13}), we conclude:

i)
The cross sections for $\rho^0$ as well as $J/\psi$ production 
are functions of $Q^2 + M^2_V$, rather than $Q^2$ 
itself. When plotted against $Q^2 + M^2_V$, one will find approximately the same 
functional behavior for the $\rho^0$ and the $J/\psi$, 
except for normalization 
differences due to $R^{(V)}$ (compare (\ref{4.2})), 
the above-mentioned factor 3/2, and  the effective value 
of $\Delta M^2_V$. In the $\rho^0$ case, $\Delta M^2_V$ corresponds to the 
level spacing, while for the $J/\psi$ the analogous quantity is determined by
the integral over the threshold factor (\ref{4.28}). 

ii) 
As soon as 
\be
Q^2 + M^2_V \gg \Lambda^2 (W^2) 
 \label{4.38} 
\ee 
the energy dependence is ``hard'', as $\Lambda^4 (W^2)$. 
This condition for the $J/\psi$ is fulfilled already for 
$Q^2$ not far from $Q^2 \ge 0$ at HERA energies, 
while $Q^2 \gg \Lambda^2 (W^2)$ is required for the $\rho^0$.  
Note that far beyond HERA energies, the increase of 
$\Lambda^2 (W^2)$ with energy will result in a soft energy 
dependence at $Q^2 \cong 0$ for the $J/\psi$ since (\ref{4.38})
will be violated, and the $J/\psi$ according to (\ref{4.37}) 
will behave as the $\rho^0$ in 
photoproduction at presently available energies. 
To stress the point again: it is relation (\ref{4.38}) 
that decides on the energy dependence. 
A $\rho^0$-like continuum state of sufficiently
high mass (consisting of approximately massless $u$ and $d$ quarks)
according to (\ref{4.6}) shows a ``hard'' energy behavior even 
in the limit of $Q^2<\Lambda^2(W^2)$.
On the other hand, a hypothetical quark of mass 
$m_q \cong 1$ GeV forming a bound state of about
2 GeV would violate (\ref{4.38}) for $Q^2 \rightarrow 0$ and sufficiently high 
energy,  
and accordingly the energy dependence would be soft.

So far we have considered the approximate evaluation of the 
quark-hadron-duality relation (\ref{4.1}).  For the
direct evaluation of (\ref{4.1})  in the massive-quark case, 
we need the generalization of (\ref{4.35}) to (\ref{4.37})
to massive quarks even beyond the threshold (\ref{4.31}).
Employing the light-cone wave functions (\ref{2.8}) and (\ref{2.9}) 
for a non-vanishing quark mass we have gone through the steps
that lead to (\ref{2.35}) to (\ref{2.37}) in the massless
case.  Upon substituting the result into (\ref{4.1}), and upon
integration over $dM^2$ from $m_q^2/z(1-z)$
to  $4 m^2_q +\Delta M_V^2$, one obtains,
\bqa
    \frac{d\sigma_{\gamma^* p \rightarrow (J/\psi , \Upsilon) p}}{dt} 
    \Bigg|_{t=0} &=& 
    \frac{\alpha R^{(J/\psi,\Upsilon)}}{32\pi^2}(\sigma^{(\infty)})^2 
     \int_{{1\over 2}(1-\overline{\Delta z})}
         ^{{1\over 2}(1+\overline{\Delta z})} dz
       \label{4.39} \\
    & &\Bigg[
         \bigl(z^2+(1-z)^2\bigr) 
         \Pi_T\Bigl(\Lambda^2(W^2),Q^2+{{m_q^2}\over{z(1-z)}},
                 M^2-{{m_q^2}\over{z(1-z)}}\Bigr)
            \nonumber \\
    &&+ \Bigl( 4Q^2z(1-z)+{{m_q^2}\over{z(1-z)}}\Bigr)\cdot   \nonumber \\
    &&~\cdot
      \Pi_L\Bigl(\Lambda^2(W^2),Q^2+{{m_q^2}\over{z(1-z)}},
                M^2-{{m_q^2}\over{z(1-z)}}\Bigr)
        \Bigg]  
     ^{4m^2_{(c,b)} + \Delta M^2_{(J/\psi,\Upsilon)}}
     _{{{m^2_{ (c,b)}}\over{z(1-z)}} }.
    \nonumber 
\eqa 
Here the functions $\Pi_T$ and $\Pi_L$ are identical to the ones
encountered earlier in the massless cases, (\ref{4.20}) and (\ref{4.18}),
respectively.  The arguments in (\ref{4.34}), however,  differ 
from the ones in
(\ref{4.20}) and (\ref{4.18}) by the substitution\footnote{Note that the 
substitution (\ref{4.40}) at the threshold (\ref{4.31}) coincides with the 
substitution (\ref{4.32}). At threshold, the massive-quark {\it cross section} is
obtained from the massless one by applying the substitution (\ref{4.40}). 
This is not true
in general, as the factor in front of $\prod_L$ in general is {\it not} 
obtained by the substitution (\ref{4.40}).}
\bqa
     Q^2 ~~~\to~~~Q^2 + {{m_q^2}\over{z(1-z)}}, \nonumber \\
     M^2 ~~~\to~~~M^2 - {{m_q^2}\over{z(1-z)}}. \label{4.40}
\eqa
The interval $\overline{\Delta z}$ in (\ref{4.39}) is related to the one
that is given in (\ref{4.25}) by the replacement of $M_{q\bar q}^2$ by 
$4m_q^2+\Delta M_V^2$, i.e.
\bq
   \overline{\Delta z} = \sqrt{{\Delta M^2_{(J/\psi,\Upsilon)}}
   \over{4m_{(c,b)}^2
        +\Delta M^2_{(J/\psi,\Upsilon)}} } \label{4.41}
\eq
For the sake of clarity, we mention that (\ref{4.39}) is to 
be understood under the constraint
\bq
    4m^2_{(c,b)}\le M^2_{(J/\psi,\Upsilon)}\le 
    4m^2_{(c,b)}+\Delta M^2_{(J/\psi,\Upsilon)},
    \label{4.42}
\eq
where $M_{(J/\psi,\Upsilon)}$ denotes the experimental value
of the vector meson mass, and 
 $\Delta M^2_{(J/\psi,\Upsilon)}$ the level-spacing, while the 
mass of charm or bottom quark, $m_{(c,b)}$, is not uniquely fixed.

We have compared the numerical result from (\ref{4.39}) with the
approximate one in (\ref{4.37}).
For $J/\psi$ production, for $Q^2 + M^2_{J/\psi} \ge 25 \, {\rm GeV}^2$, the
approximation (\ref{4.37}) overestimates the result from (\ref{4.39})
by less than 10\%. For the case of $\Upsilon$ production, for $Q^2 \cong 0$, for
later reference we note that the approximation result is about 40\% larger than
the result of the exact evaluation. 

In order to compare the cross sections for $J/\psi$ and $\Upsilon$ production
with the cross sections for $\rho^0$ production, it is useful to remove the 
effect due to differences in the quark content and to define an 
``enhancement factor'' $E^{(V)}$ by 
\bq
      E^{(V)} 
    \equiv {{R^{(\rho^0)}}\over {R^{(V)}}}
              {{\sigma_{\gamma^* p\rightarrow Vp}}\over
                 {\sigma^{\gamma^* p \rightarrow \rho^0 p} }},~~~~~
       V=J/\psi,\Upsilon,
\label{4.43}
\eq
where the cross sections are to be evaluated at identical fixed values of 
$Q^2 + M^2_V$ for the different vector mesons and at the same energy $W$. 
For $Q^2 +M^2_V$ sufficiently large, the slope parameters of the 
$t$-distribution, $b (Q^2  + M^2_V)$, 
are experimentally known to become identical,
$b(Q^2 + M^2_V \gsim 30 \, {\rm GeV}^2) \cong 4.5 \, {\rm GeV}^{-2}$, and 
(\ref{4.43})
may be evaluated by inserting forward-production cross sections. 
Inserting the approximations (\ref{4.37}) and (\ref{4.10}), we have 
\bq
   E^{(V)}_{\rm appr.} = \frac{3}{2}
\frac{\Delta F^2 (m^2_q, \Delta M^2_V)}{\Delta M^2_
{\rho^0}} \cdot \frac{1}{\left( 1 + \frac{\Lambda^2 (W^2)}{Q^2 + M_V^2}\right)^2},
     ~~~~~V=J/\psi,\Upsilon.
  \label{4.44}
\eq
The dependence on $Q^2 + M^2_V$ leads to some increase of $E^{(J/\psi)}$
with $Q^2 + M^2_V$ that is confirmed by the more reliable evaluation of 
(\ref{4.43}) based on (\ref{4.39}) and (\ref{4.15}). 
The enhancement (\ref{4.43}) and (\ref{4.44}) is recognized as a genuine 
massive-quark effect with respect to the factor 3/2 following from 
$4m^2_q \cong M^2_V$, and the threshold factor $\Delta F^2 (m^2_q , 
\Delta M^2_V)$ that replaces $\Delta M^2_V$ in the massive-quark case. 

\begin{table}
\begin{tabular}{|c | c|c|c|c|c|c|}\hline
 $V$ & $R^{(\rho^0)}/R^{(V)}$ & $M_V$ (GeV) & $\Delta M^2_V ({\rm GeV}^2)$ &
$m_q$ (GeV) & $E^{(V)}_{\rm appr.}$ & 
$E^{(V)}$ \\
\hline
$\rho^0$ & 1 & 0.77 & 1.0 & 0 & --- & --- \\
$J / \psi$ & $\frac{9}{8}$ & 3.096 & 4.0 & 1.5 & 2.01 & 2.25 \\ 
$\Upsilon$ & $ \frac{9}{2}$ & 9.460 & 11.0 & 4.6 & 3.32 &2.69 \\
\hline
\end{tabular}
\caption{ We show the input parameters used in the calculations  
of $\rho^0 , J/\psi $ and $\Upsilon$ production, as well as  the
enhancement of $J / \psi$ and $\Upsilon$ production relative to
$\rho^0$ production in the 
approximate evaluation and in the exact numerical 
evaluation of quark-hadron duality. The enhancements of $J/\Psi$ 
and $\Upsilon$ refer to identical values of 
$Q^2+M_V^2=89.48$ GeV$^2$.} 
\end{table}

We have numerically evaluated (\ref{4.43}), assuming a universal slope 
parameter $b$ for 
$\rho^0 , J / \psi$ and $\Upsilon$. Inserting  (\ref{4.39})
and (\ref{4.15}), we find the enhancement factor of  
Table 1 at $W=90$ GeV and $Q^2 + M^2_V = 89.48 \, {\rm GeV}^2$
that is relevant for $\Upsilon$ photoproduction. 
In Table 1, we also present the results obtained from the 
approximation (\ref{4.44}). 
The deviation from the results from (\ref{4.43}) for the $J / \psi$ is
due to the mentioned excess of about 10\% of the approximation (\ref{4.10})
for the $\rho^0$ meson with respect to the exact result. The larger 
deviation in the case of the $\Upsilon$ is largely due to the above 
mentioned 40\% deviation of (\ref{4.37}) for the $\Upsilon$.

\subsection{Comparison with experiment}

In this Section we compare the $Q^2$ dependence, the $W$ dependence and the 
absolute normalization of the  vector-meson-production 
cross sections with the experimental data. 

From (\ref{4.10}) and (\ref{4.37}) the $Q^2$ dependence of 
vector-meson production at 
$Q^2 + M^2_V \gg \Lambda^2 (W^2)$ is determined by $(Q^2 + M^2_V)^{-n}$
where $n=3$. 
A fit to the experimental data \cite{Janssen}
for $\rho^0$ production including data from $Q^2 \cong 10 \, {\rm GeV}^2$ 
to $Q^2 \cong 50 \, {\rm GeV}^2$ lead to $n_{\rho^0} = 2.60 \pm 0.04$. 
For $J/\psi$ production, a fit in the interval $12 \, {\rm GeV}^2 \le Q^2 + 
M^2_{J/\psi} \le 60 \, {\rm GeV}^2$ by the ZEUS collaboration gave 
$n_{J/\psi} = 2.72 \pm 0.10$.\cite{Tandler} 
Both, $n_{\rho^0}$ and $n_{J/\psi}$ are 
consistent with the prediction of $n=3$, taking into account the fairly low 
value of the lower end of the interval in $Q^2 + M^2_V$ used 
in the fits.  It only exceeds 
$\Lambda^2 (W\cong 70 GeV) \cong 3.5 \, {\rm GeV}^2$ 
by a factor of 3. 

Both, the H1 and ZEUS collaborations have fitted the energy dependence of 
their data by a simple power law, 
\be
\sigma_{\gamma^*p \rightarrow V p} (W^2 , Q^2) \sim W^{\delta^{(V)} 
(Q^2)} .
\label{4.45}
\ee
We have compared our theoretical energy dependence to the power law 
(\ref{4.45}) by adapting our energy dependence to this power-law form 
in a restricted energy range (chosen as in the experiments) via 
\be
\delta^{(V)} (Q^2 + M^2_V) = \frac{1}{\ln \frac{W_2}{W_1}} \ln 
\frac{\sigma_{\gamma^* p \rightarrow Vp} (W_2 , Q^2 + M^2_V)}{\sigma_
{\gamma^*p\rightarrow Vp} (W_1, Q^2 + M^2_V)} .
\label{4.46}
\ee
The consistency of our prediction with the experimental data \cite{HERA-II,jpsiphoto}
for $\rho^0$ and $J/\psi$ 
production is shown in fig.6.

We turn to the absolute normalization of the vector-meson-production 
cross section. According to (\ref{4.10}) and (\ref{4.13}) 
as well as (\ref{4.37})
the normalization is first of all determined by the 
product of $R^{(V)} \cdot (\sigma^{(\infty)})^2$, where $R^{(V)}$ 
according to (\ref{4.2}) contains 
the charges of the relevant quark flavors, and $\sigma^{(\infty)}$ denotes 
the asymptotic value of the (universal) color-dipole
 cross section (\ref{2.21}). 
In Table 1, according to (\ref{4.2}), we 
show  the frequently used ratio $R^{(\rho^0)} / R^{(V)}$
that normalizes the cross sections to the $\rho^0$ case 
as far as the quark content of the vector mesons is concerned. 
For the universal dipole cross section, 
$\sigma^{(\infty)}$, we use
\be
\sigma^{(\infty)} = 68 \, {\rm GeV}^{-2} = 27.5 \, mb . 
\label{4.47}
\ee
This value is consistent with the value from the analysis 
of $\sigma_{\gamma^* p}(W^2,Q^2)$, compare (\ref{2.34}). 
In addition to $R^{(V)}$ and $\sigma^{(\infty)}$, 
the normalization of the cross
sections in (\ref{4.10}), (\ref{4.13}) and (\ref{4.37}) 
is determined by the integration interval 
$\Delta M^2_V$ entering via
quark-hadron duality as well as the mass of 
the respective quark. The interval 
$\Delta M^2_V$ follows from the vector-meson level spacing, 
and for the 
$\rho^0$, in the approximate treatment (\ref{4.10}) and (\ref{4.13}), 
$\Delta M^2_{\rho^0} = 1 \, {\rm GeV}^2$
directly multiplies the diffractive production cross 
section at $M^2 \equiv M^2_{\rho^0}$. 
In the massive quark case, the integrated threshold factor 
$\Delta F^2(m_q^2,\Delta M_V^2)$ from (\ref{4.28}),
effectively replaces $\Delta M^2_V$. 
In Table 1, we have collected all relevant quantities, 
including the quark masses. 

Finally, the experimental cross sections include an integration over
the (exponential) $t$ dependence, $\exp (bt)$, that implies a 
factor of $1/b$. We note that\cite{Janssen, kreisel} 
\bqa
   b_{\rho^0} (Q^2 + M^2_{\rho} \cong M_{J/\psi}^2) &\cong& 5.5 \,{\rm GeV}^{-2} , \nonumber \\
   b_{J/\psi} (Q^2 + M^2_{J/\psi} \cong M_{J/\psi}^2) &\cong& 4.5 \,{\rm GeV}^{-2}, 
\label{4.48}
\eqa
while for lower values of $Q^2 + M^2_V$,  $b_{\rho^0}$ increases to 
\bq
   b_{\rho^0} (Q^2 + M^2_{\rho^0} \cong 1 {\rm GeV}^2 ) 
   \cong 7.5 \, {\rm GeV}^{-2} . 
\label{4.49}
\eq

With these preparations, it is a simple matter to discuss the relative
normalization of $\rho^0 , J/\psi$ and $\Upsilon$ production. 
 For the ratio of $J/\Psi$ photoproduction ($Q^2=0$) to 
$\rho^0$ production
at $Q^2+M_{\rho^0}^2=M_{J/\Psi}^2$, from (\ref{4.15}), (\ref{4.39})
and (\ref{4.48}), we find
\be 
\frac{9}{8} \sigma^{(J/\psi)} / \sigma^{(\rho^0)} = 1.42,
\label{4.50}
\ee
while for $\Upsilon$ photoproduction, from Table 1 
\be 
     E^{(\Upsilon)} =
\frac{9}{2} \sigma^{(\Upsilon)} / \sigma^{(\rho^0)} = 2.69.
\label{4.51}
\ee
The factors (\ref{4.50}) and (\ref{4.51}) are consistent 
with the enhancements found experimentally for $J/\psi$ \cite{levy}
as well as $\Upsilon$ production \cite{upsilon}. Note that a replacement of 
the integrated threshold 
factor (\ref{4.28}) by the level spacings would have led 
to drastically increased enhancements.

In figs. 7 to 10, we show a comparison of our predictions with the 
experimental 
$Q^2$ and $W$ dependence of $\rho^0$ and $J/\psi$ production. 
The theoretical results shown in the figures are based on the use of (\ref{4.1})
without further appproximation, i.e. they are based on
(\ref{4.15}) for the $\rho^0$ and on (\ref{4.39})
for the $J/\psi$.
For $\rho^0$ production, for fig.7 
we inserted a constant slope parameter 
\be 
b_{\rho^0} = 7.5 \, {\rm GeV}^{-2} , 
\label{4.52}
\ee
as well as the $Q^2$-dependent slope \cite{kreisel}
from (\ref{3.15}).
In fig.8, 
we show the $W$ dependence of $\rho^0$ production 
at various values of $Q^2$
compared with data from ZEUS and H1 measurements \cite{HERA-II}.
In fig.9, 
we show the longitudinal-to-transverse ratio for $\rho^0$ production. 
According to figs.7 to 9, we have satisfactory agreement for the
$Q^2$ and $W$ dependence, including normalization, as well as
the longitudinal-to-transverse ratio.
In fig.10 we show the comparison with photoproduction of $J/\psi$ mesons.
The agreement in the $W$ dependence is very satisfactory indeed. 

We conclude that our two-gluon-exchange QCD-based color-dipole approach
(called GVD-CDP) yields a parameter-free representation 
(except for using the experimental value of the slope parameter $b$) 
of vector-meson production. 
The dependence on the variable $Q^2 + M^2_V$ 
is a strict consequence of including the quark mass of 
magnitude $4m^2_q \cong M^2_V$ into the
lightcone wave function of the photon. The effective value of the gluon 
transverse momentum, $\Lambda^2 (W^2)$, sets the scale for the large,
$Q^2 + M^2_V \gg \Lambda^2 (W^2)$, and small,
$Q^2 + M^2_V \ll \Lambda^2 (W^2)$, 
regimes  with associated strong and weak $W$ dependences. 
The relative normalizations of the cross sections 
are a genuine consequence of the quark masses relative to the 
vector-meson masses; for the $\rho^0$ the approximation of massless quarks is
relevant, while for the $J /\psi$ and $\Upsilon$ 
we have $M^2_{J /\psi} \cong
4m^2_c$ and $M^2_\Upsilon \cong 4 m^2_b$, respectively.

\subsection{A brief reference to the literature on vector-meson photo-
and electroproduction}

The first theoretical papers on electroproduction of vector mesons
\cite{Fraas, Cho}
some thirtyfour years ago were based on (simple diagonal) vector-meson 
dominance with $s$-channel helicity conservation. From the coupling of 
the vector meson to a conserved source as  required by electromagnetic current
conservation, it was concluded that production by longitudinal photons 
should dominate the cross section via $R_{L/T} \sim Q^2 / M^2_V$ - a prediction
that has stood the test of time and also appears in our present paper. 
The unpolarized cross section, however, was predicted \cite{Fraas, Cho} 
to only decrease as $1 / Q^2$ in strong disagreement with present-day
experimental results that require an asymptotic behavior approximately as
$1 / Q^6$. A revival \cite{sss1} of off-diagonal GVD \cite{Read}
implied a $1 / Q^4$ dependence for the unpolarized cross section that in 
fact was found \cite{sss1}
to be consistent with the experimental data up to around 
$Q^2 \le 20 \, {\rm GeV}^2$ available at the time, including the ratio of 
$R_{L/T}$. Since the traditional vector-dominance 
approach relies on (soft) Pomeron exchange, it is unable to 
describe the strong $W$ dependence observed for 
sufficiently large values of $Q^2 + M^2_V$. 

The modern QCD theory of vector-meson production uses the notion of the 
Pomeron as a two-gluon-exchange object \cite{Low}
throughout. Various approaches differ in their application of methods from 
pQCD \cite{Ryskin,Brodsky}, pQCD combined with quark-hadron duality 
\cite{Levin}, 
the use of non-perturbative QCD approaches \cite{dl1,Cudell,Dosch1}
and combinations of both \cite{dgs,Don}.
The pQCD approach, valid at sufficiently high $Q^2$ led to an asymptotic 
$(1/Q^6) \alpha_s (Q^2) xg (x, Q^2)$ behavior of the dominant longitudinal 
cross section \cite{Brodsky},
closely related to our asymptotic behavior as $(1 /Q^6) 
\sigma^{(\infty)} \Lambda^2 (W^2)$.
In its spirit more closely related to our 
approach are the investigations in ref. \cite{Levin}
in so far as they make use of quark-hadron duality, however in conjunction
with conventional parameterizations of the gluon structure function. 

The approach of the present paper has the distinctive feature of 
being applicable at all $Q^2 \ge 0$ for light and heavy vector mesons. The 
increase of the effective value of the gluon transverse momentum with energy, 
$\Lambda^2(W^2)$,
determined from the DIS measurements of the total photoabsorption 
and expressed in terms of three fit parameters 
is sufficient to yield a parameter-free
and unambiguous 
description of (forward) vector meson production. The final expression for 
the cross sections are simple and transparent, and they put the 
(approximate) universal 
dependence on $Q^2 + M^2_V$, as well as the $W$ dependence and the relative 
normalization of the cross sections for the production of different 
vector mesons on a firm footing.

\section{Conclusion}

After many years of experimental and theoretical efforts, it seems that 
a coherent picture of DIS in the low $x$ diffraction region, 
DVCS, and vector-meson production has emerged. The $Q^2$ dependence and 
the relative normalization of the cross sections is well understood from our
QCD-based approach of the GVD-CDP.
Also the scale for the remarkable transition from a soft to a hard 
energy dependence in DIS, DVCS and 
vector meson production, as well as the very occurrence of this transition, 
is understood, even though an ab initio 
prediction of the power of $W$ responsible for the 
increase with energy, $W$, is beyond the
scope of our present understanding.

\newpage

\begin{figure}[htbp]\centering
\epsfysize=5cm
\centering{\epsffile{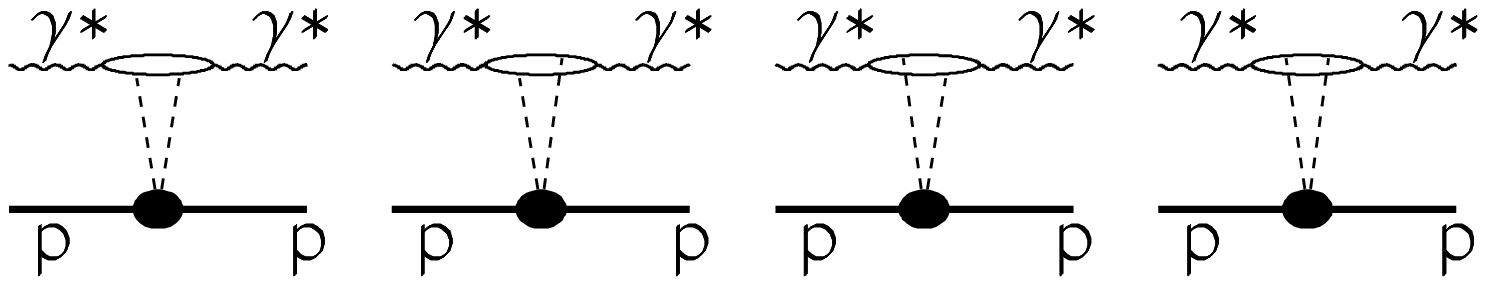}}
\caption{The forward Compton amplitude.}
\label{fig1}
\end{figure}

\begin{figure}[htbp]\centering
\epsfysize=5cm
\centering{\epsffile{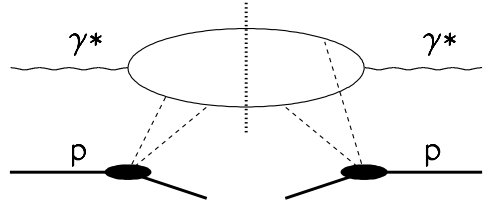}}
\caption{One of the 16 diagrams for diffractive production.  
The vertical line indicates the unitarity cut corresponding to the
diffractively produced final states, $(q\bar q)^J$.  
Production of (discrete or continuum) vector states corresponds to 
$(q\bar q)^J$ production with $J=1$.}
\label{fig2}
\end{figure}

\newpage

\begin{figure}[htbp]\centering
\epsfysize=13cm
\centering{\epsffile{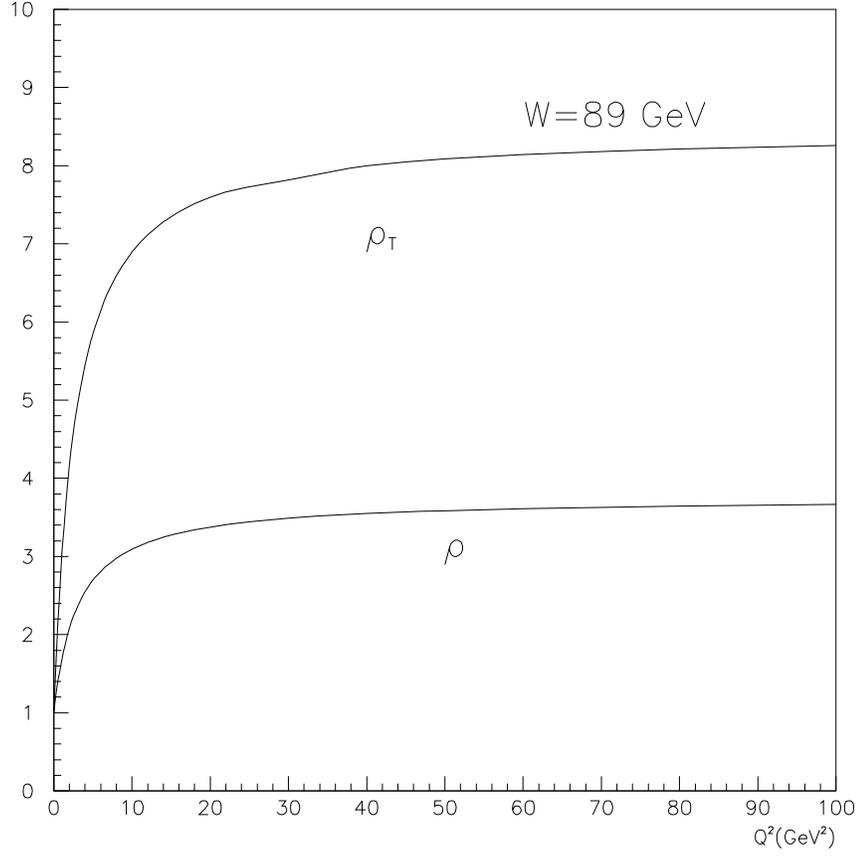}}
\caption{The ratio $\rho_T$ is calculated using the transverse part of
the total photoabsorption cross section,
$\rho_T \equiv 16\pi d\sigma(\gamma^*p \to \gamma p)/dt\Big/
\sigma^2_{\gamma^*_T p}(W^2,Q^2)$, while $\rho$ uses
$\sigma_{\gamma^* p}\equiv \sigma_{\gamma^*_T p}+ \sigma_{\gamma^*_L p}$,
i.e. $\rho \equiv 16\pi d\sigma(\gamma^*p \to \gamma p)/dt\Big/
\sigma^2_{\gamma^* p}(W^2,Q^2)$.}
\label{fig3}
\end{figure}

\vfill\eject

\begin{figure}[htbp]\centering
\epsfysize=13cm
\centering{\epsffile{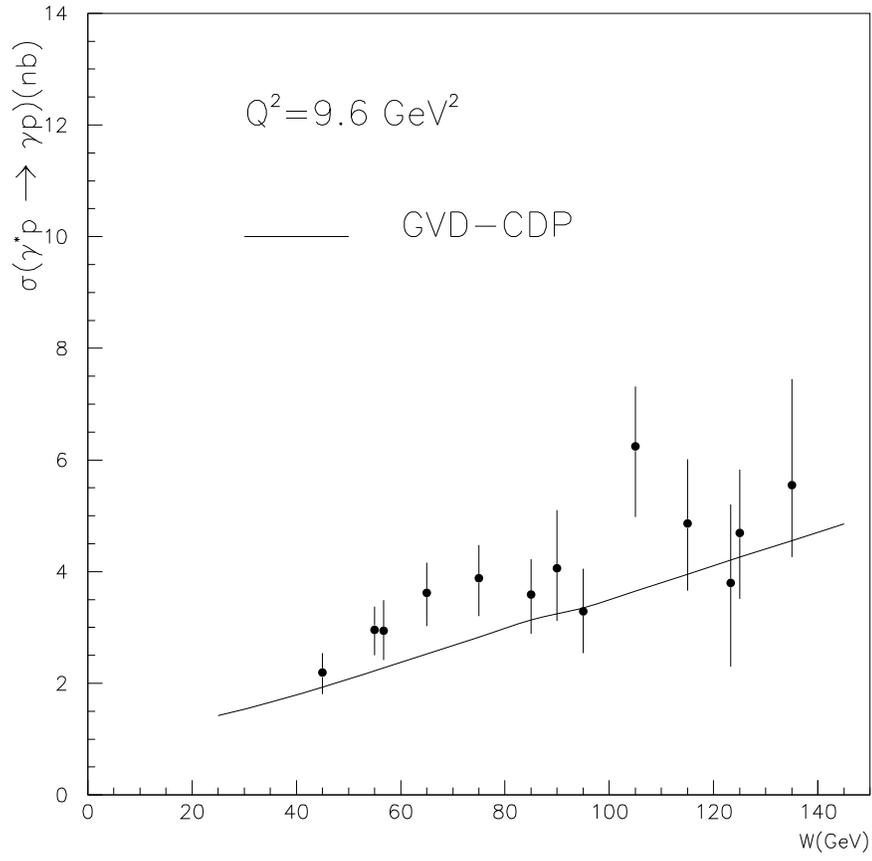}}
\caption{The $W$-dependence for DVCS for $Q^2 = 9.6~GeV^2$ compared
with the prediction from the QCD-based GVD-CDP.}
\label{fig4}
\end{figure}
\vfill\eject

\begin{figure}[htbp]\centering
\epsfysize=13cm
\centering{\epsffile{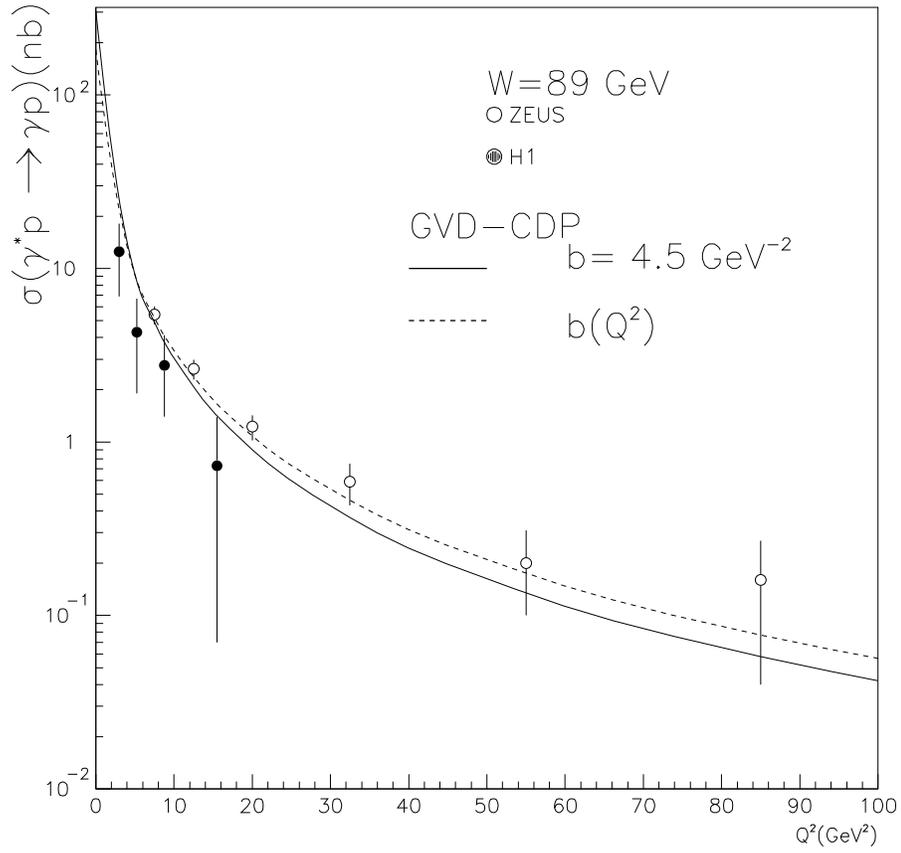}}
\caption{As fig. 4, but for the $Q^2$ dependence at fixed $W = 89~GeV$.
The H1 data have been appropriately shifted \cite{HERA-I} 
in energy from the 
H1 value of $W=$ 75 GeV to $W=$ 89 GeV.}
\label{fig5}
\end{figure}
\vfill\eject

\begin{figure}[htbp]\centering
\epsfysize=13cm
\centering{\epsffile{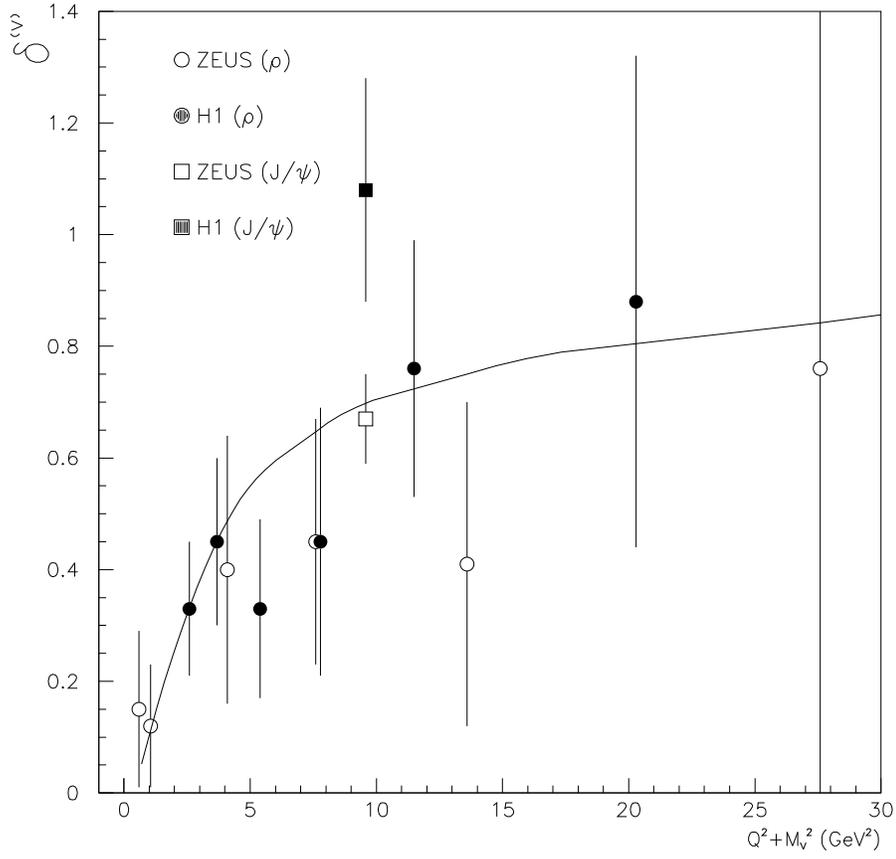}}
\caption{The exponent $\delta^{(V)}$ in a parameterization of the 
energy-dependence of the experimental cross section by 
$W^{\delta^{(V)} (Q^2 + M^2_V)}$ compared
with the predictions from the QCD-based GVD-CDP.}
\label{fig6}
\end{figure}
\vfill\eject

\begin{figure}[htbp]\centering
\epsfysize=13cm
\centering{\epsffile{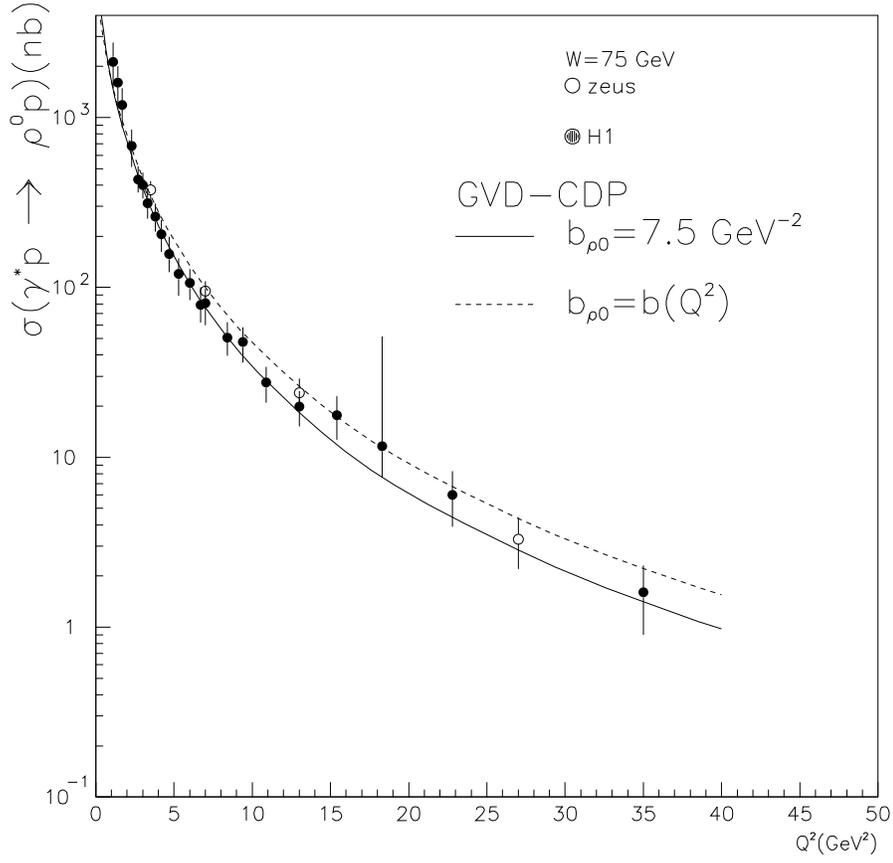}}
\caption{The $Q^2$ dependence of $\rho^0$ production, $\gamma^*p \to
\rho^0 p$, at fixed $W = 75~GeV$, compared with the predictions from
the QCD-based GVD-CDP for $b_{\rho_0}=7.5$ GeV$^{-2}$ and
a $Q^2$-dependent slope, $b_{\rho^0}(Q^2)$ from (\ref{3.15}).}
\label{fig7}
\end{figure}
\vfill\eject

\begin{figure}[htbp]\centering
\epsfysize=10cm
\centering{\epsffile{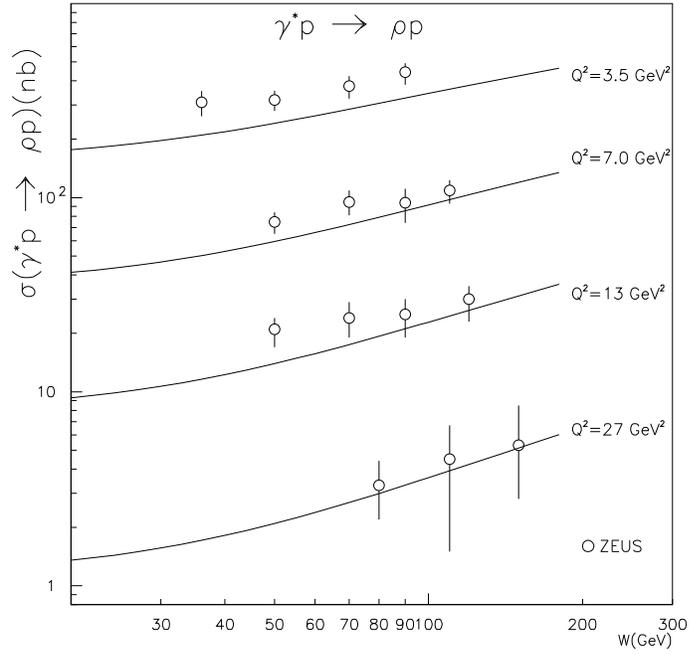}}
\centering{\epsffile{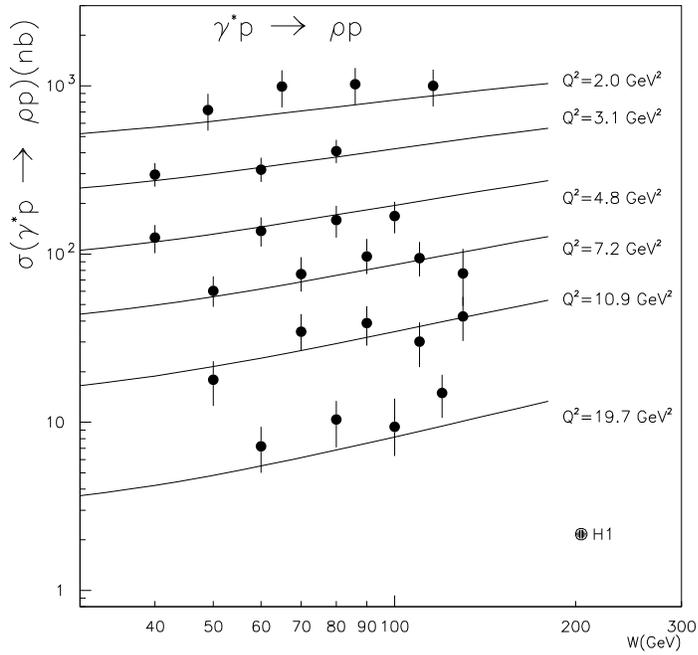}}
\caption{The $W$ dependence of $\rho^0$ production for various values
of $Q^2$ compared with the QCD-based GVD-CDP.}
\label{fig8}
\end{figure}
\vfill\eject

\begin{figure}[htbp]\centering
\epsfysize=13cm
\centering{\epsffile{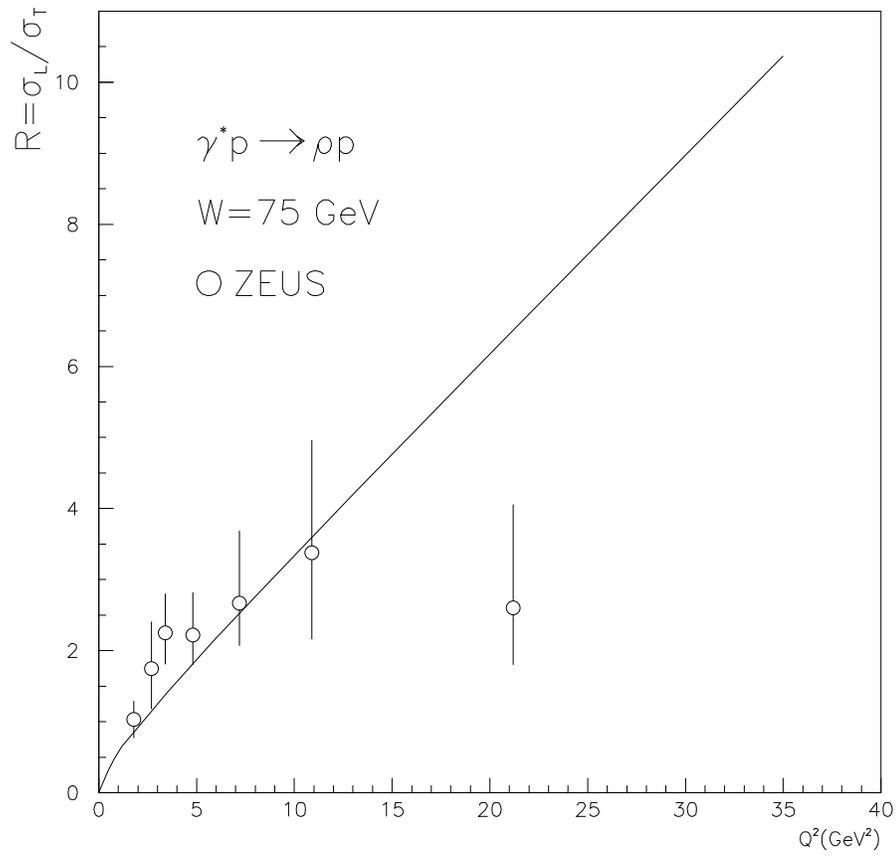}}
\caption{The longitudinal to transverse ratio, $R_{L/T}$ for $\gamma^* p
\to \rho^0 p$ as a function of $Q^2$.}
\label{fig9}
\end{figure}
\vfill\eject

\begin{figure}[htbp]\centering
\epsfysize=13cm
\centering{\epsffile{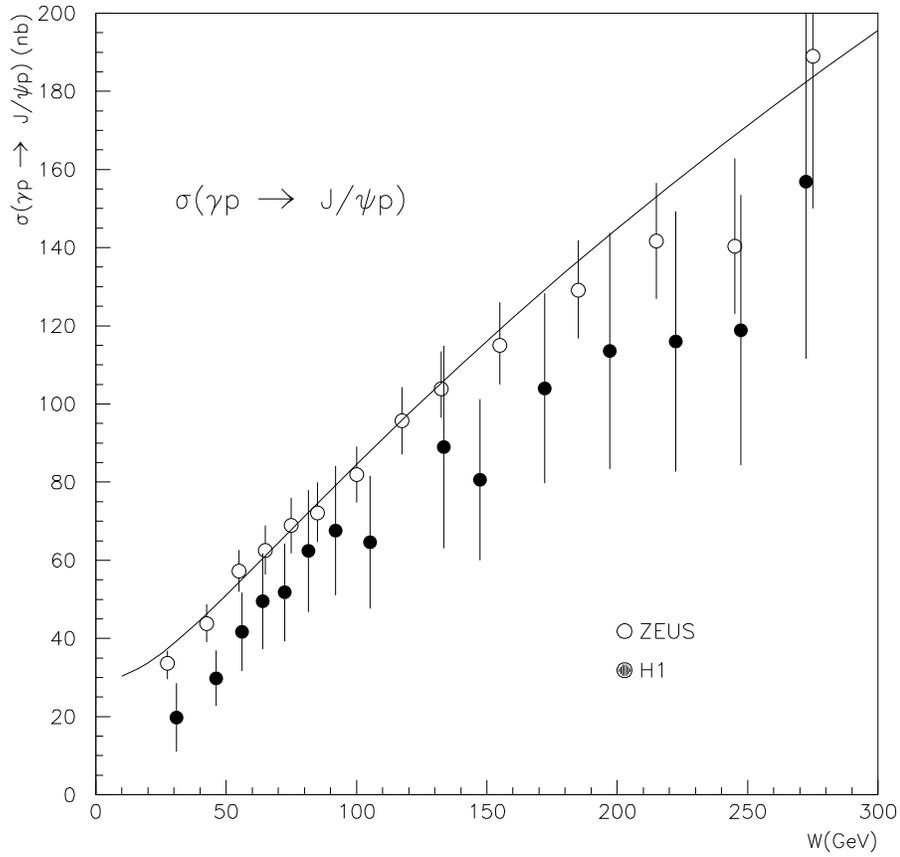}}
\caption{The $W$ dependence of the $J/\psi$ photoproduction
cross section, compared with the theoretical prediction from the QCD-based
GVD-CDP. }
\label{fig10}
\end{figure}
\vfill\eject

\end{document}